\documentclass[journal]{IEEEtran}
\usepackage{subfigure}
\usepackage{graphicx}
\usepackage{epstopdf}
\usepackage{epsfig}
\usepackage{psfrag}
\usepackage{amssymb}
\usepackage{color}
\usepackage{algorithm}
\usepackage[noend]{algpseudocode}
\usepackage{balance}

\usepackage{url}
\usepackage{amsmath}
\usepackage{fancyhdr}
\usepackage{multirow}

\usepackage[utf8]{inputenc}
\usepackage{amsmath}

\begin{document}

\title{Characterizing Task Completion Latencies in Fog Computing} 
\author{Maria Gorlatova, \emph{Member, IEEE}, Hazer Inaltekin, \emph{Member, IEEE}, Mung Chiang, \emph{Fellow, IEEE}.\thanks{M. Gorlatova is with the Department of Electrical and Computer
Engineering, Duke University, Durham, NC 27708, USA (e-mail:
maria.gorlatova@duke.edu). H. Inaltekin is with the Department of Electrical and Electronic Engineering,
University of Melbourne, Parkville, VIC 3010, Australia (e-mail:
hazeri@unimelb.edu.au). M. Chiang is with the School of Electrical and Computer Engineering, Purdue
University, West Lafayette, IN 47907, USA (e-mail: chiang@purdue.edu).}}

\maketitle 

\begin{abstract}
Fog computing, which distributes computing resources to multiple locations between the Internet of Things (IoT) devices and the cloud, is attracting considerable attention from academia and industry.
Yet, despite the excitement about the potential of fog computing, few comprehensive quantitative characteristics of the properties of fog computing architectures have been conducted. 
In this paper we examine the properties of task completion latencies in fog computing. First, we present the results of our empirical benchmarking-based study of task completion latencies. 
The study covered a range of settings, and uniquely considered both traditional and serverless fog computing execution points. It demonstrated the range of execution point characteristics in different locations and the relative stability of latency characteristics for a given location. It also highlighted properties of serverless execution that are not incorporated in existing fog computing algorithms. Second, we present a framework we developed for co-optimizing task completion quality and latency, which was inspired by the insights of our empirical study. 
We describe fog computing task assignment problems we formulated under this framework, and present the algorithms we developed for solving them. 
\end{abstract}

\section{Introduction}

\par 
Fog computing is an emerging paradigm that distributes computing resources
closer to the end users along the ``cloud-to-things'' continuum~\cite{OpenFogReferenceArchitecture2017}.
Fog computing is receiving increasing attention from industry and academia alike~\cite{ETSI_MEC2017,AmazonGreenGrass,LambdaAtEdge2017,AzureEdge2017,Xiao2017QoE,Tan2017Online,Souza2016Towards,Jia2017Optimal}
due in part to its potential for enabling advances in Internet of Things (IoT) applications. Yet, despite the broad attention that fog computing is receiving, and despite many recent developments in this field~\cite{AmazonGreenGrass,AzureEdge2017,OpenFogReferenceArchitecture2017}, comprehensive quantitative characterizations of core properties of fog computing architectures are currently lacking. Our work addresses a part of this gap, as we describe below. 

\par
\begin{figure}[t]
  \centering
   \includegraphics[width=0.9\linewidth]{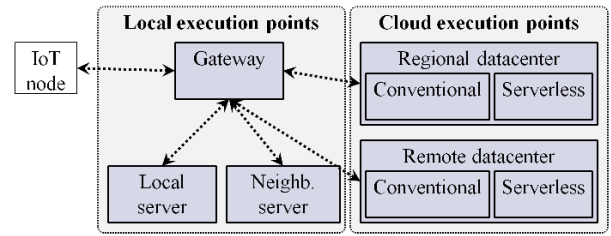}
    \vspace{-0.3cm}
    \caption{A multi-execution-point fog computing system serving responsive IoT applications. 
  \label{fig:OverallSettings}}
\end{figure}

In this work we focus on characterizing properties of task completion latencies in fog computing, for systems with heterogeneous, geographically distributed task execution points, 
as shown in Fig.~\ref{fig:OverallSettings}. While observations about differences in latencies in different settings have recently been made~\cite{Li2010CloudComp,Gao2015Cloudlets,Chen2017Empirical,Perez2017Experimental}, to the best of our knowledge \emph{we are the first to comprehensively examine many latency properties important for designing and developing fog task assignment algorithms, such as latency models, variability, and stability}. 

We focus on characterizing task completion latencies for responsive IoT applications, in which IoT nodes rely on fog computing for timely responses to their requests -- as opposed to, for example, fog-based execution of data analytics tasks without explicit deadlines. 
The need for timely responses arises in both human-facing and machine-to-machine applications -- e.g., in gaming, augmented and virtual reality~\cite{Chu2016Towards}, user guidance and feedback~\cite{Georgiev2016LEO,Ra2011Odessa,Chen2017Empirical}, motion control in cars, autonomous drones, and industrial machinery~\cite{Qi2016Design,Liberatore2006Integrated}, and control of parameters in cyber-physical systems. 
Responsive applications are considered to be one of the primary use cases for fog computing~\cite{OpenFogReferenceArchitecture2017,ETSI_MEC2017}. 

Our work takes steps towards using fog computing systems to support a range of task execution latency preferences. 
As both human-facing and machine-to-machine interactions need to be supported by fog systems, 
it is important to develop mechanisms for handling diverse response timelines and time utility functions. 
Considering a wide range of time sensitivities is additionally important for scaling and automating future IoT deployments. While in small-scale deployments it may be possible to tune the responsive behavior of an IoT application manually on a per-application basis, the continuing expansion of the IoT will necessitate automated management of latency-related application preferences and requirements. 


Below we first provide an overview of task execution points we include in our study, 
then describe our contributions. 

\subsection{Execution points in fog computing \label{Sect:Architecture}}

\begin{table}[t]
\centering
  \caption{Example computing capabilities of different fog computing execution points.~\label{table:DeviceCapabilities}
  }
  \vspace{-0.3cm}
  \begin{tabular}{|p{0.17\linewidth} |p{0.2\linewidth} | p{0.26\linewidth} |p{0.2\linewidth}|}
   \hline
  Fog comput. point & Device or service & CPU, \# cores, freq. & Additional comp. capab. \\ \hline
  Gateway & Raspberry Pi 3 Model B~\cite{RaspPiModelB} & 4@1.2 GHz  & GPU 
   \\ \hline
  Local server 
  & PowerEdge R930~\cite{PowerEdgeR930} & \hbox{4--24}@\hbox{2.2--3.5} GHz & GPUs \\ \hline 
  Cloud, conventional & AWS C2~\cite{EC2InstanceTypes} & \hbox{1--128}@\hbox{2.3--3.3} GHz & Can add GPUs, FPGAs\\ \hline
  Cloud, serverless & AWS Lambda~\cite{AmazonLambda} & Fractional parts of AWS C2 cores & None \\ \hline
  \end{tabular} 
\end{table} 

\par
Execution points present in fog computing systems have different properties, as we describe below. Several representative options of different fog node types are shown in Table~\ref{table:DeviceCapabilities}.

\textbf{Gateways}: Most modern IoT architectures connect low-end IoT devices to the Internet
    via gateways placed nearby. 
    These gateways can be stationary, such as smart home gateways~\cite{SmartThings} and car dashboard nodes~\cite{CARDuino}, or mobile, such as mobile phones and tablets. 
    The gateways are \emph{consumer electronics-grade} -- they are notably less computationally capable than enterprize-grade servers. For example, as can be seen in Table~\ref{table:DeviceCapabilities}, the CPU speed of a Raspberry Pi-based gateway is 1/2 of the speed of a small server. 
    We include a gateway-grade local node in our study. 
 
\textbf{Local servers}:
    Despite the rising popularity of cloud computing services,
    on-premise computing remains an essential feature of many organizations, including universities.
    Ranging from small stand-alone machines to full on-site datacenters,
    local computing infrastructure offers computing options in relative proximity to the local IoT nodes, for example, in-building, or on the same campus~\cite{streiffer2017eprivateeye}. 
    We include a local server-grade node in our study. 

\textbf{Cloud, conventional}: Traditional cloud services' computing capabilities are located in one of several global datacenters (e.g., AWS offers services in 16 regions, Google Cloud -- in 7).
    Cloud computing has been commercially available for over a decade. As part of the cloud services, multiple options with different underlying hardware and different performance guarantees are currently available~\cite{EC2InstanceTypes}.
    We include two AWS execution points, located in different regions, in our study. 
    
\textbf{Cloud, serverless}: 
    \emph{Serverless} cloud computing allows running computing using cloud datacenter resources without managing specific computing instances: rather than being provisioned ahead of time, individual functions are \emph{instantiated on demand} when they are requested.
    First introduced by Amazon in 2014, serverless computing is now offered by all major cloud providers~\cite{GoogleCloudFunctions,AmazonLambda,AzureFunctions,OpenWhisk}, and is actively promoted specifically for IoT applications~\cite{AmazonLambda,AzureFunctions}. 
    We include two AWS-based serverless execution points and one Microsoft Azure-based serverless execution point in our study. 
    To the best of our knowledge, we are \emph{the first to closely examine the properties of serverless computing as an execution option for latency-sensitive tasks in fog computing}, and \emph{the first to incorporate properties of serverless execution points in our problem formulations and algorithms}. 
    
\subsection{Our contributions}


We closely examine statistical properties of task execution latencies in heterogeneous fog computing systems. 

First, to examine task completion latencies in fog computing, we developed and deployed a set of mini-benchmarks in 6 different locations, which include local nodes of different grades, conventional cloud computing services in two different regions, and serverless computing options from both AWS and Microsoft Azure. Using this experimental infrastructure, we conducted a set of targetted experiments with an edge node invoking these benchmarks from different locations and in different conditions, including from 10 different environments on a trip between the regions, and continuously for 30 days with different inter-invocation times. Altogether we obtained over 1,000 hours of measurements. 

Our experimental study elucidated several important properties of task execution latencies, including the heterogeneity of their properties across different locations, and their relative stability with respect to time. The study also demonstrated properties of serverless execution options not examined before. Based on the results of our study, we argue for the need to empirically learn statistical properties of task completion latencies in fog deployments. We demonstrate that appropriate latency characterizations can be obtained relatively quickly. 

Next, inspired by the observed task completion latency properties, we developed a framework that co-optimizes time and quality of task execution in fog computing. \emph{Unlike existing frameworks, this framework applies to cases where different execution points' latencies have different statistical properties} (which we observed in our experiments), and \emph{allows operating over different task timeliness preferences}. We applied this framework to the problem of assigning tasks to fog execution points. Our problem formulations and the algorithms we developed take into account properties of fog computing systems not examined before. Our numerical results 
demonstrate that the developed framework correctly operates over diverse task timeliness requirements. 

\par 
The rest of the paper is organized as follows. 
Section~\ref{Sect:RelatedWork} describes the related work. Section~\ref{Sect:Measurements} presents the setup we developed for our measurement study. 
Sections~\ref{Sect:OnTheTrip}--\ref{sect:ServerlessExecutionNotes} describe our insights into variability and stability of task completion latencies, and into properties of serverless execution. Section~\ref{Sect:Model} presents our framework for task quality-latency co-optimization. 
Section~\ref{Application:TaskAssignment} demonstrates an application of this framework to task assignments in fog computing. Section~\ref{Sect:Conclusions} concludes the paper. 

\section{Related work \label{Sect:RelatedWork}}

Empirical observations of task completion latencies in fog computing have been made in~\cite{Li2010CloudComp,Gao2015Cloudlets,Chen2017Empirical,Perez2017Experimental}; 
such studies typically focus on first-order latency characterizations, often in context of a specific application optimized for fog execution. By contrast, we conduct a larger-scale study, focusing on task completion latency models, variability, stability, and other complex properties. 
To the best of our knowledge, we are \emph{the first to examine serverless execution alongside other fog computing options} in fog latency characterizations. Observations about complex nature of latency in serverless computing have recently been made in~\cite{McGrath2017,KeepingLambdasWarm2017}; these studies did not model the latency, and did not examine its implications on fog computing algorithms.    

Our framework for co-optimizing task completion latency and quality, 
which optimizes task utilities with an intrinsic quality and a time-dependent component, 
is inspired by the lines of work on time-dependent utilities~\cite{RealTimeSystemsBook,Sun2017Update,ju2012symphoney}, multi-quality approaches~\cite{Han2016MCDNN,ran2018deepdecision}, and time-quality co-optimization techniques~\cite{zilberstein1996using}, conducted independently in different fields. 
Time-dependent utilities appear in real-time systems~\cite{RealTimeSystemsBook},
explorations of sensor data ``age of information''~\cite{Sun2017Update},
and in models of human preferences in mobile systems~\cite{ju2012symphoney}. Optimizing task execution quality levels has been examined in the fields of approximate computing~\cite{Mittal2016Survey,Miguel2016Anytime}, and is starting to be considered as a possibility in mobile edge computing~\cite{Han2016MCDNN,ran2018deepdecision}. Joint consideration of intrinsic and time-dependent utilities has been proposed in the field of anytime algorithms~\cite{zilberstein1996using} -- this approach inspired ours. 

Finally, task assignments in fog computing systems have been examined in multiple lines of work. Such approaches typically aim to minimize task completion delays~\cite{Souza2016Towards,Jia2017Optimal,Xiao2017QoE,Tan2017Online}. 
By contrast, in this work we co-optimize task completion quality and latency. We further comment on the difference between the traditional approaches and ours in Section~\ref{Sect:CoOptimizationTimeQuality}. Additionally, to the best of our knowledge, our problem formulation and our algorithms uniquely take into account the potential co-existence of capacitated and infinite-capacity nodes in fog computing systems.  


\section{Characterizing Fog Systems Latencies\label{Sect:Measurements}}

First, in Sect.~\ref{Sect:OurExperimentalSetup} we describe the experimental setup we developed for our latency measurements. Then in Sect.~\ref{Sect:LocalNodes} we comment on the expected task latency tradeoffs we observed. Last, in Sect.~\ref{Sect:ResponseTimesAdvanced} we describe the fit of the Generalized Extreme Value (G.E.V.) probability distributions to the observed task completion latencies. 

\subsection{Experimental tasks and execution points\label{Sect:OurExperimentalSetup}}
\begin{table}[t]
\centering
\caption{Fog computing execution points we deployed.~\label{table:DevicesInExperiments}}
  \vspace{-0.3cm}
 \begin{tabular}{|p{0.02\linewidth}|p{0.27\linewidth} | p{0.23\linewidth} | p{0.28\linewidth}|}
  \hline
  \# & Comp. node & Locations & CPU specifications \\
 \hline
 1 & Local server-grade node & Campus & Intel Xeon E5-2609 4@2.4 GHz \\ \hline
 2 & Local consumer-grade node & Residential location & Intel Atom 230 1@1.60GHz \\ \hline
 3 & AWS C2 t2.micro inst. & Reg. 1 (closest), Reg. 2 (remote) & Intel Xeon E5-2676 1@2.40GHz \\ \hline
 4 & AWS Lambda & Reg. 1 (closest), Reg. 2 (remote) & Various, from AWS EC2 pool 
 \\ \hline
 5 & Microsoft Azure serverless funct. & Reg. 1 (closest) & Various, from Microsoft Azure pool \\ \hline
 \end{tabular} 
\end{table}

\begin{table}[t]
\centering
  \caption{Mini-benchmarks we deployed.~\label{table:TasksWeRun}}
    \vspace{-0.3cm}
  \begin{tabular}{|p{0.02\linewidth}|p{0.22\linewidth} | p{0.37\linewidth} | p{0.2\linewidth}|}
  \hline
  \# & Task & Description & Exec. options \\
  \hline
  1 & $\pi$ Calculations (PIC) & Calculate $\pi$ to many decimal places via Leibnitz approximation. & 5,000 -- 500,000 iter.\\ \hline 
  2 & Process Stored File (PSF) & Calculate statistics on a dataset read from node memory. & 500 -- 50,000 lines read \\ \hline
  3 & File send-and-process (FSP) & Calculate statistics on a dataset sent to the node. & 500 -- 10,000 lines transm.\\ \hline
  \end{tabular}
\end{table}

\par We deployed mini-benchmarks on fog computing points listed in Table~\ref{table:DevicesInExperiments}. These points include both local nodes and cloud services, and both conventional and serverless cloud execution options. 

\par 
The mini-benchmarks we developed (in Python) and deployed on the above-listed nodes are summarized in Table~\ref{table:TasksWeRun}. These mini-benchmarks stress different aspects of a computing system. The \emph{$\pi$ Calculations (PIC)} task stresses the CPU (a similar Super $\pi$ procedure is used to characterize the performance of conventional and overclocked CPUs~\cite{SuperPi2017}). The \emph{Process Stored File (PSF)} task stresses memory access. The \emph{File Send-and-Process (FSP)} task stresses the communications. The benchmarks are executed with a range of execution options shown in Table~\ref{table:TasksWeRun}. The benchmarks do not correspond to one specific application, but rather are representative of CPU-limited, memory-limited, and communications-limited fog computing tasks.  

\par 
On conventional computing points we deployed the benchmarks over the popular Flask web services framework~\cite{FlaskFramework}, with communication over the HTTP request-response protocol. 
The codebase deployed in all conventional computing nodes is identical. For serverless execution, which relies on proprietary deployment agents, we adapted our code slightly to adhere to the required serverless input and output handling. 

\par 
As local execution points we selected a server-grade option (\#1 in Table~\ref{table:DevicesInExperiments}) with an Intel Xeon CPU, and a consumer-grade option (\#2 in Table~\ref{table:DevicesInExperiments}) with an Intel Atom CPU. 
These execution points are not fully dedicated to our purposes, but are lightly loaded. We included different classes of local devices to correspond to a variety of local execution options potentially available in fog computing architectures, as described in Section~\ref{Sect:Architecture}. 

\par
As cloud execution points we used two AWS datacenters, one in our geographic region (US East, North Virginia, $\approx$300 miles from our campus; Region~1 in Table~\ref{table:DevicesInExperiments}), and one that is far away (US West, Oregon, $>2,500$ miles from campus; Region~2 in Table~\ref{table:DevicesInExperiments}). For conventional cloud-based execution we used EC2 t2.micro instances~\cite{EC2InstanceTypes}. 
We also included serverless AWS ( AWS Lambda)~\cite{AmazonLambda} and Microsoft Azure options~\cite{AzureFunctions} 
in our experiments (\#4 and \#5 in Table~\ref{table:DevicesInExperiments}). We ran serverless Microsoft Azure functions in Region~1, and serverless AWS options in the same Region~1 and~2 computing centers as the conventional cloud services.

\begin{table}[t]
\centering
  \caption{ICMP ping times for fog nodes accessed from an on-campus location and a nearby residential location.~\label{table:PingTimes}}
    \vspace{-0.3cm}
  \begin{tabular}{|p{0.2\linewidth}|p{0.15\linewidth}| p{0.08\linewidth} | p{0.08\linewidth} |p{0.08\linewidth} |p{0.08\linewidth}|}
  \hline
   & \multicolumn{5}{|c|}{ICMP ping latency, ms, for different fog nodes} \\ \cline{2-6}
   & Meas. & Camp. & Resid. & Reg.~1 & Reg.~2\\
  \hline
  On-campus  & Median & 2.3 & 16 & 8.7 & 70.4 \\ \cline{2-6}
   edge node & 90$^{\textrm{th}}$ perc. & 4.4 & 21.4 & 9.6 & 71.4 \\ \hline \hline
  Residential & Median & 15.2 & 0.9 & 18.7 & 86.8 \\ \cline{2-6}
   edge node & 90$^{\textrm{th}}$ perc.& 21.1 & 1.4 & 23.7 & 92.1 \\ \hline
  \end{tabular}
\end{table} 

\par
The majority of our experiments were conducted with an edge node, represented by a wirelessly connected laptop, calling the deployed services from either an on-campus location or a close-by residential location ($<5$ miles from campus). 
The ICMP ping response times for the different fog nodes accessed from these location are shown in Table~\ref{table:PingTimes}. As expected, at each location a local fog node can be reached faster than the closest cloud node (2.3~ms vs.~8.7~ms for the on-campus location, 0.9~ms vs.~18.7~ms for the residential one). We also performed several experiments with other locations (e.g., the experiments described in Sect.~\ref{Sect:OnTheTrip}), and several with the edge node connected via Ethernet 
(e.g., the experiments described in Sect.~\ref{Sect:ResponseTimeStability}). 

\par 
We measured task completion latencies in their entirety, as experienced by nodes accessing fog platform and receiving a reply from it, with the incurred delays including communication, service initialization, and task execution components. To partially decouple the impacts of local and remote latency impacts, we 
interleaved the invocations of benchmarks on different execution points, local and remote. In particular, this approach helped us distinguish between the WiFi-associated local connectivity disruptions, which affect communications to all execution points, and the conditions affecting individual execution points. 

\begin{table}[t]
\centering
 \caption{Median and 95$^{\textrm{th}}$ percentile task completion latencies for the PIC benchmark requested by an edge node located on campus. Each data point in this table is based on over 1,200 data samples. 
  ~\label{table:DelaysPi}}
  \vspace{-0.3cm}
  \begin{tabular}{|p{0.15\linewidth}|| p{0.065\linewidth} | p{0.08\linewidth} | p{0.095\linewidth}||p{0.065\linewidth}| p{0.08\linewidth}|p{0.095\linewidth}|}
  \hline
  Exec. & \multicolumn{3}{|c||}{Median latency (s)} & \multicolumn{3}{|c|}{95$^{\textrm{th}}$ perc.~latency (s)}
  \\ \cline{2-7}
  point & 5,000 iter. & 50,000 iter. & 500,000 iter. & 5,000 iter. & 50,000 iter. & 500,000 iter.\\
  \hline
  Server & 0.02 & 0.06 & 0.47  & 0.04 & 0.08 & 0.64 \\ \hline
  Cons. node & 0.08 & 0.38 & 3.0 & 0.12 & 0.61 & 3.5 \\ \hline
  AWS EC2, Reg.~1 & 0.03  & 0.05 & 0.37 & 0.04 & 0.07 & 0.52\\ \hline
  AWS EC2, Reg.~2 & 0.15  & 0.18  & 0.48 & 0.16 & 0.19 & 0.62 \\ \hline
  AWS Lambda, Reg.~1 & 0.46 & 0.52  & 1.19 &0.57& 0.65& 1.37 \\ \hline
  AWS Lambda, Reg.~2 & 0.72 & 0.79 & 1.05 & 0.86 & 0.92 & 1.27\\ \hline
  \end{tabular}

\end{table}

Altogether we collected over 1,000 hours of experimental data, which included several multi-day experiments, some lasting as long as 30 days. Example summaries of the measured task completion latencies, for the PIC benchmark executed on different fog nodes with different execution options, are shown in Table~\ref{table:DelaysPi}. Each value in this table is based on more than 1,200 individual data samples. 

\subsection{Known and intuitive tradeoffs observed~\label{Sect:LocalNodes}}

\begin{figure}[t]
\centering
\subfigure[PIC, EC2 vs.~local server.]{
\includegraphics[width = 0.45\linewidth]{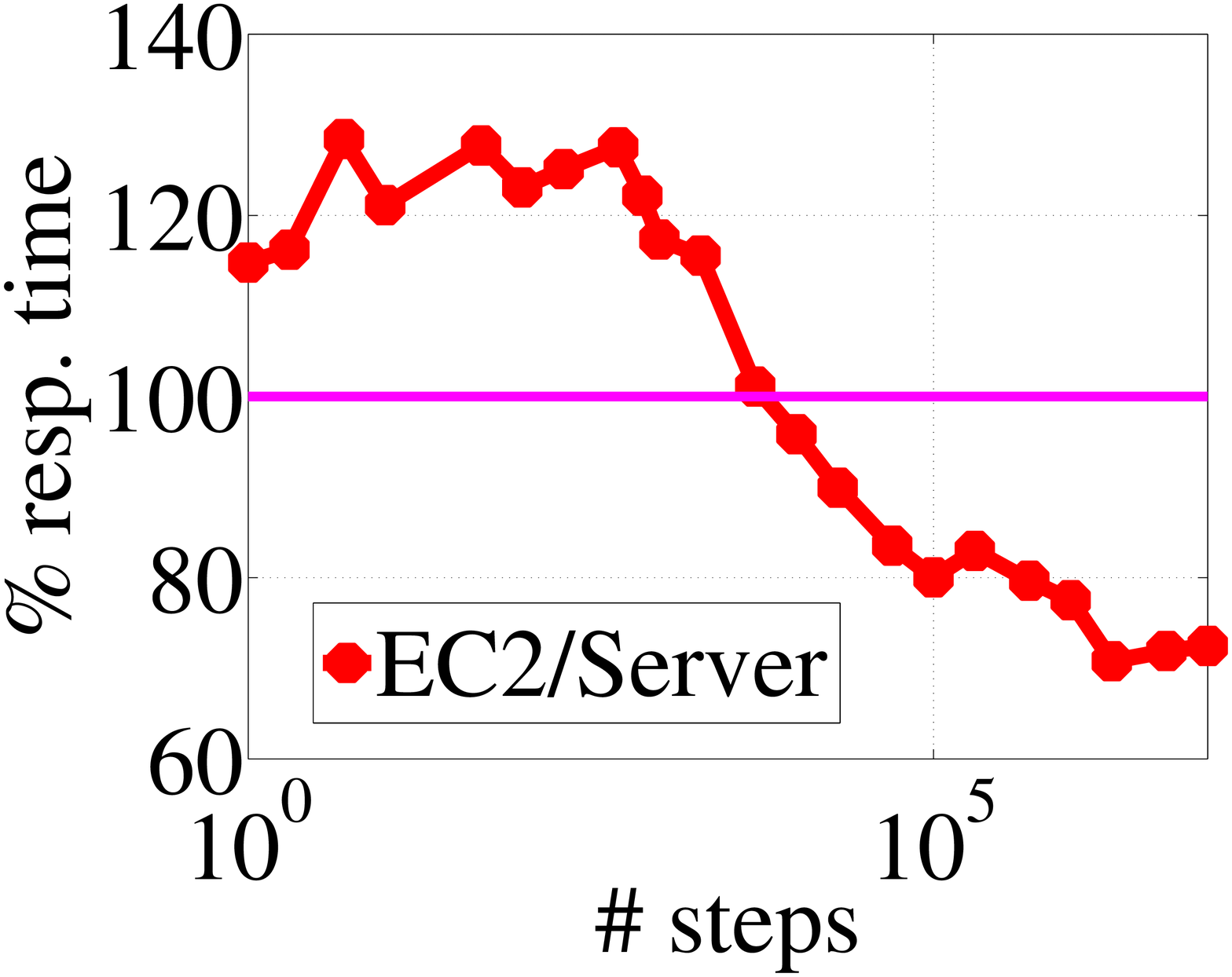} \label{fig:executionTimesPi}}
\subfigure[FSP, EC2 vs.~local consumer-grade node.]{
\includegraphics[width = 0.45\linewidth]{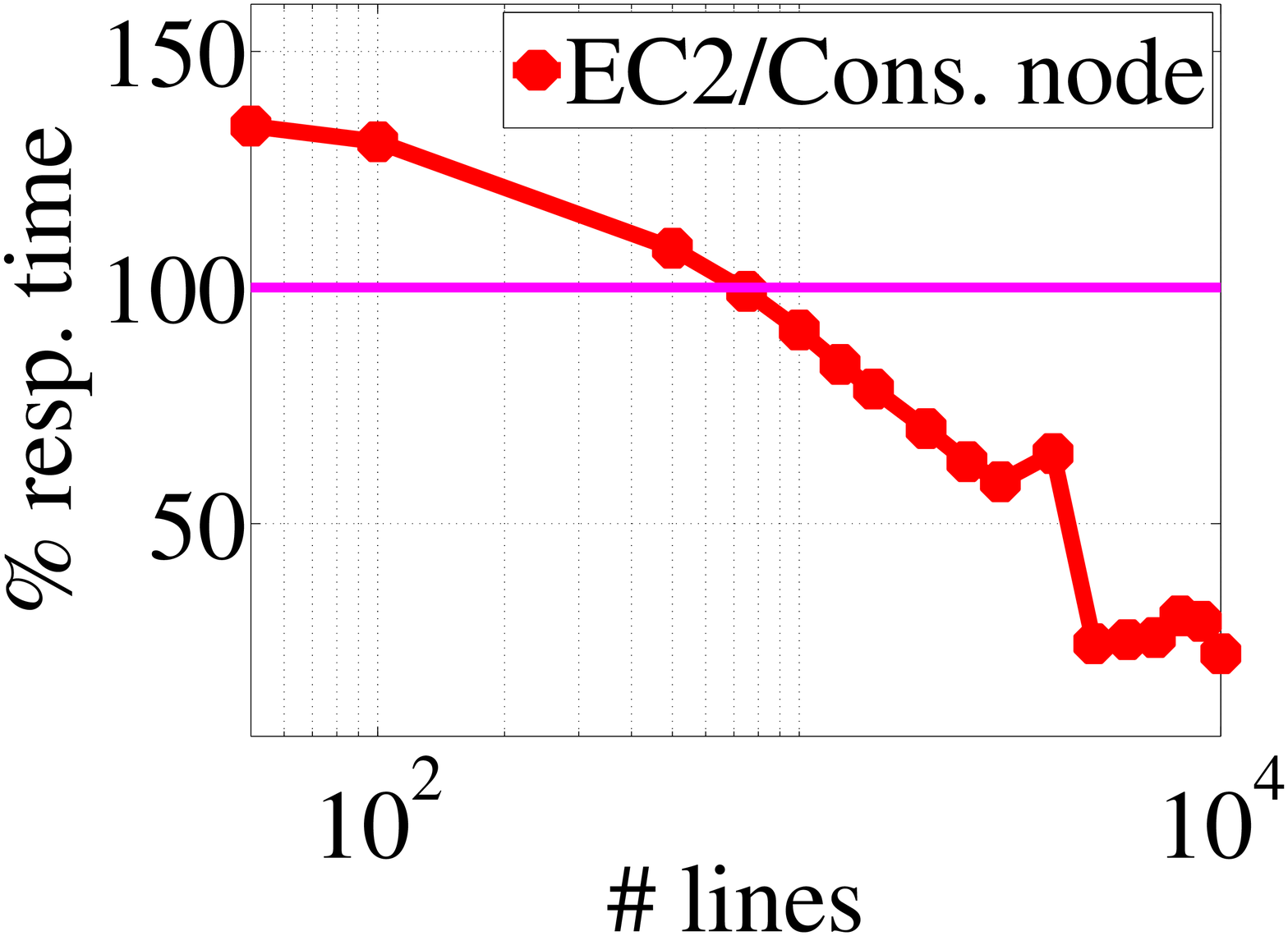} \label{fig:ConsumerNodeTask3}}
\vspace{-0.3cm}
\caption{AWS EC2 Region 1 task completion latencies, as \% of local node task completion latencies, for different task sizes. Local nodes are faster than the cloud nodes up to a certain level of task complexity.~\label{fig:executionTimeDifferencesWork}}
\end{figure} 

\par
Our measurements demonstrate that, with both communication and computing taken into consideration, task completion latency is smaller for local nodes than for
the conventional cloud nodes, up to a certain complexity of execution. 
This tradeoff is intuitive: 
for small tasks, the latency is dominated by the communication time, which is smaller for local nodes, while for larger tasks, it is dominated by the execution time, which can be smaller for more capable cloud nodes.
The tradeoff can be observed, for example, in Fig.~\ref{fig:executionTimeDifferencesWork}, which demonstrates, 
for the PIC and FSP tasks, the conventional cloud point latency as a percentage of the local computing point latency: up to a particular level of complexity of a task, local execution of these tasks completes faster than the cloud execution.  

\par
Another intuitive observation we make is that the tradeoffs between local and cloud-based execution can be vastly different for consumer-grade and server-grade local nodes. As task complexity increases, the performance of consumer-grade nodes substantially lags the performance of server-grade nodes (see the 3$^{\textrm{rd}}$ and the 6$^{\textrm{th}}$ columns in Table~\ref{table:DelaysPi}, for example).
Correspondingly, while server-grade ``cloudlets'' and the cloud may offer similar performance, in fog computing systems with consumer-grade gateways, the best execution point for a particular task needs to be carefully chosen to optimize the overall system performance. 
This observation also alludes to the need to differentiate consumer-grade and server-grade nodes in fog \emph{resource pooling}, as pooling consumer-grade nodes may not be advantageous when the fog system serves high-complexity tasks, or when local server-grade cloudlets are available. 

\par
Finally, we also observed that \emph{connections with the cloud are notably faster on-campus than off-campus} (e.g., as listed in Table~\ref{table:PingTimes}, we measured ICMP ping delays as 8.7~ms vs. 18.7~ms for Region~1, and 70.4~ms vs. 86.8~ms for Region~2). Unusually low latency of campus-to-cloud connections has been noted before~\cite{Gao2015Cloudlets}; we observed it on two different university campuses. This observation suggests that \emph{on-campus experiments do not fully capture the experience of the majority of the users of fog computing architectures}, and suggests the need to include off-campus locations in experimental examinations of fog computing architectures.

\subsection{Modeling task completion latencies\label{Sect:ResponseTimesAdvanced}}
\begin{figure}[t]
\centering
\subfigure[AWS EC2.]{
\includegraphics[width = 0.45\linewidth]{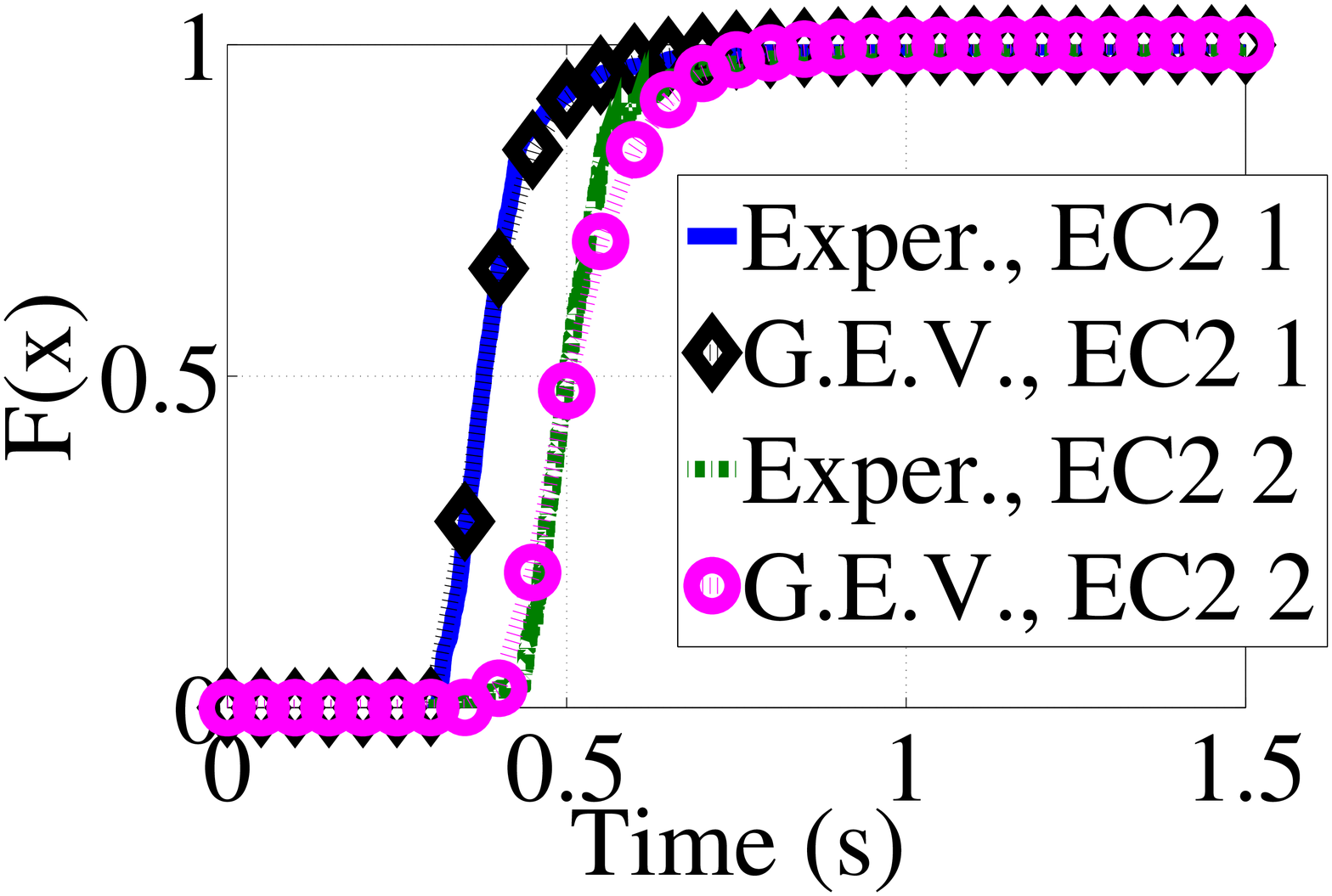} \label{fig:AWSLambdaEastFitted}
}
\subfigure[AWS Lambda (``$\Lambda$'').]{
\includegraphics[width = 0.45\linewidth]{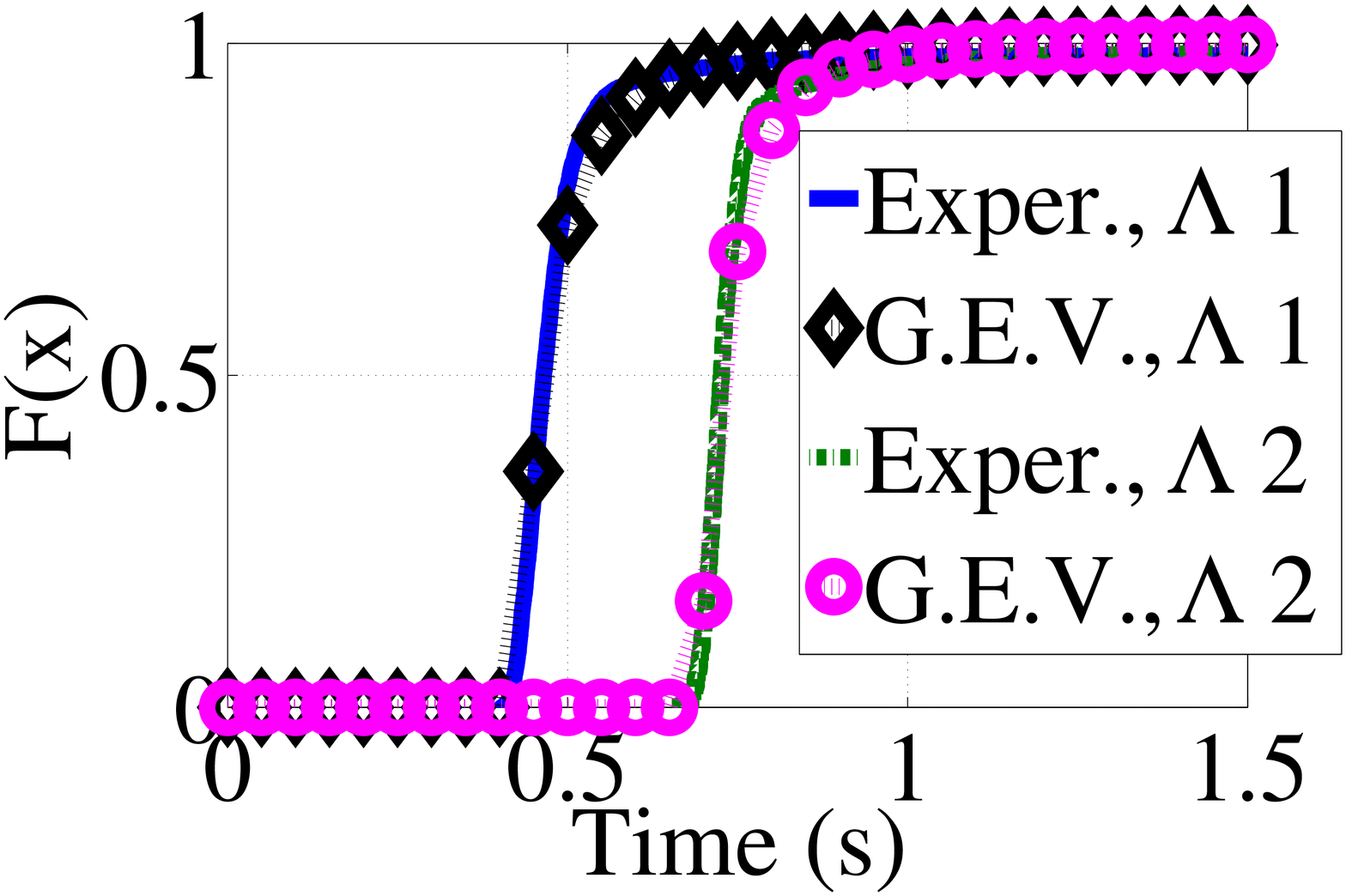} \label{fig:AWSLambdaEastFitted}
}
\vspace{-0.25cm}
\caption{The CDFs of measured latencies for several settings, shown with the best-fitting Generalized Extreme Value (G.E.V.) distributions.\label{fig:ResponsesFitted}}
\vspace{-0.3cm}
\end{figure} 

\begin{table}[t!]
\centering
  \caption{G.E.V. distributions' fit to the experimental latency data, for some of the examined benchmarks.~\label{table:QualityFits}}
    \vspace{-0.3cm}
  \begin{tabular}{|p{0.3\linewidth}|p{0.07\linewidth}|p{0.15\linewidth} | p{0.3\linewidth}|}
  \hline
 Settings & $N$ & Distr. support (s) & G.E.V.~parameters \\
  \hline 
 PIC, campus, node 1, 500,000 iter. & 1444 &  0.42 -- 5.8 & Shape 0.34, scale 0.04, loc 0.48 \\ \hline
 PSF, campus, node 3, 5,000 lines & 3961 &  0.15 -- 3.2 & Shape 0.35, scale 0.004, loc. 0.16 \\ \hline
 FSP, residential, node 1, 10,000 lines & 2710 & 0.09 -- 0.64 & Shape 0.12, scale 0.008, loc. 0.11 \\ 
\hline
  \end{tabular}
\end{table}

\par
To better understand the properties of task completion latencies, we fitted the experimentally obtained task completion latency samples to a set of 20 common statistical distributions using conventional maximum likelihood estimation techniques.\footnote{The full set of distributions is: Beta, Binominal, Birnbaum-Saunders, Exponential, Extreme Value, Gamma, Generalized Extreme Value, Generalized Pareto, Inverse Gaussian, Logistic, Log-logistic, Lognormal, Nakagami, Negative Binomial, Normal, Poisson, Rayleigh, Rician, $t$ Location-scale, Weibull.} 

\par We observe that in the majority of the cases we examine, experimental latency values match well to Generalized Extreme Value (G.E.V.) distributions with a positive shape parameter, that is, a Type II G.E.V.~distribution, which are also known as Frechet or Inverse Weibull. For example, Fig.~\ref{fig:ResponsesFitted} shows several experimentally obtained CDFs and the CDFs of fitted G.E.V.~distributions, while Table~\ref{table:QualityFits} shows the parameters of 
the fitted G.E.V.~distributions for three other experiments for which the G.E.V.~is the best fitting distribution of all examined. 
Inverse Weibull distributions have previously been observed to fit HTTP server processing times~\cite{Cao2004Stochastic}; as the underlying communication mechanism in our experiments is HTTP, it is logical that these distributions would be a fit for some experiments, especially those where HTTP-related mechanisms dominate task completion latencies. 

\par
We observe, also, a close fit of G.E.V.~distributions to serverless execution points' latencies, as can be seen in Fig.~\ref{fig:ResponsesFitted}(b), for example. 
G.E.V.~distributions typically arise in the study of order statistics. 
We hypothesize that internal 
cloud providers' service provisioning mechanisms behind serverless operations involve a minimization or a maximization of a performance quality, and thus are readily described by the G.E.V.~distributions.

\par 
Our observations are limited to relatively simple tasks we use in our benchmarking. 
Complex operations with runtimes that vary widely for different inputs (e.g., cascading classifiers with multiple exit points~\cite{TeerapittayanonDeep2017}, 
or complex interactive applications with many components~\cite{Chen2017Empirical}), 
may have task completion latencies defined by algorithm-related, rather than infrastructure-related, phenomena. We do, however, expect relatively simple tasks to be an important part of the IoT tasks served by fog architectures.

\section{Task Completion Latencies In Different Local Conditions\label{Sect:OnTheTrip}}

\begin{table}[t]
\centering
\caption{Task completion latencies captured, for execution points in Regions 1 and 2, 
in 10 different environments.~\label{table:TripToSeattle}}
  \vspace{-0.3cm}
 \begin{tabular}{|p{0.02\linewidth}|p{0.14\linewidth} | p{0.33\linewidth} | p{0.33\linewidth}|}
  \hline
  \# & Loc. & Exec. point: AWS EC2, Reg. 1 & Exec. point: AWS EC2, Reg. 2 \\
 \hline
 1 & Campus, Reg.~1 & $m$=0.34, $p_{10}$=0.31, $p_{90}$=0.41, $sp$=0.1  ($N$=721) & $m$=0.48, $p_{10}$=0.43, $p_{90}$=0.53, $sp$=0.1 ($N$=293)  \\ \hline
 2 & Resid., Reg.~1 & $m$=0.53, $p_{10}$=0.44, $p_{90}$=0.61, $sp$=0.17 ($N$=8627) & $m$=0.495, $p_{10}$=0.49, $p_{90}$=0.92, $sp$=0.43 ($N$=2814) \\ \hline
 3 & Airport, Reg.~1 
 & $m$=0.40, $p_{10}$=0.35, $p_{90}$=0.46, $sp$=0.11 ($N$=303) 
 & $m$=0.50, $p_{10}$=0.46, $p_{90}$=0.57, $sp$=0.11 ($N$=91)\\ \hline
 4 & Flight, Reg.~1 to mid. & $m$=0.56, $p_{10}$=0.47, $p_{90}$=0.89, $sp$=0.42 ($N$=721)  & $m$=0.66, $p_{10}$=0.55, $p_{90}$=0.98, $sp$=0.42 ($N$=254)
 \\ \hline
 5 & Airport, middle & $m$=0.43, $p_{10}$=0.36, $p_{90}$=0.79, $sp$=0.43 ($N$=624) & $m$=0.55, $p_{10}$=0.47, $p_{90}$=1.46, $sp$=0.99 ($N$=177) \\ \hline
 6 & Flight, mid. to Reg.~2 & $m$=1.92, $p_{10}$=1.79, $p_{90}$=2.19, $sp$=0.39 ($N$=1768) & $m$=1.84, $p_{10}$=1.71, $p_{90}$=2.11, $sp$=0.39 ($N$=589) \\ \hline
 7 & Hotel, Reg.~2 & 
$m$=0.44, $p_{10}$=0.44, $p_{90}$=0.51, $sp$=0.07
 ($N$=8510) &  
$m$=0.33, $p_{10}$=0.33, $p_{90}$=0.40, $sp$=0.07
 ($N$=2906) \\\hline
 8 & Campus, Reg.~2  & 
$m$=0.51, $p_{10}$=0.43, $p_{90}$=0.63, $sp$=0.20 ($N$=1504)
 & 
$m$=0.38, $p_{10}$=0.30, $p_{90}$=0.49, $sp$=0.20 ($N$=501)
\\ \hline
 9 & Airport, Reg.~2 & $m$=0.55, $p_{10}$=0.43, $p_{90}$=0.75, $sp$=0.32 ($N$=766)  & 
$m$=0.42, $p_{10}$=0.31, $p_{90}$=0.60, $sp$=0.30 ($N$=249) 
\\ \hline
 10 & Flight, Reg.~2 to mid. & 
 $m$=4.40, $p_{10}$=4.01, $p_{90}$=4.77, $sp$=0.77 ($N$=426) & 
$m$=4.28, $p_{10}$=3.93, $p_{90}$=4.71, $sp$=0.78 ($N$=161)
\\ \hline
\end{tabular}
\end{table}

\begin{figure}[t]
\centering
\subfigure[7 environments in Reg.~1, Reg.~2, and in-between the regions.]{
\includegraphics[width = 0.45\linewidth]{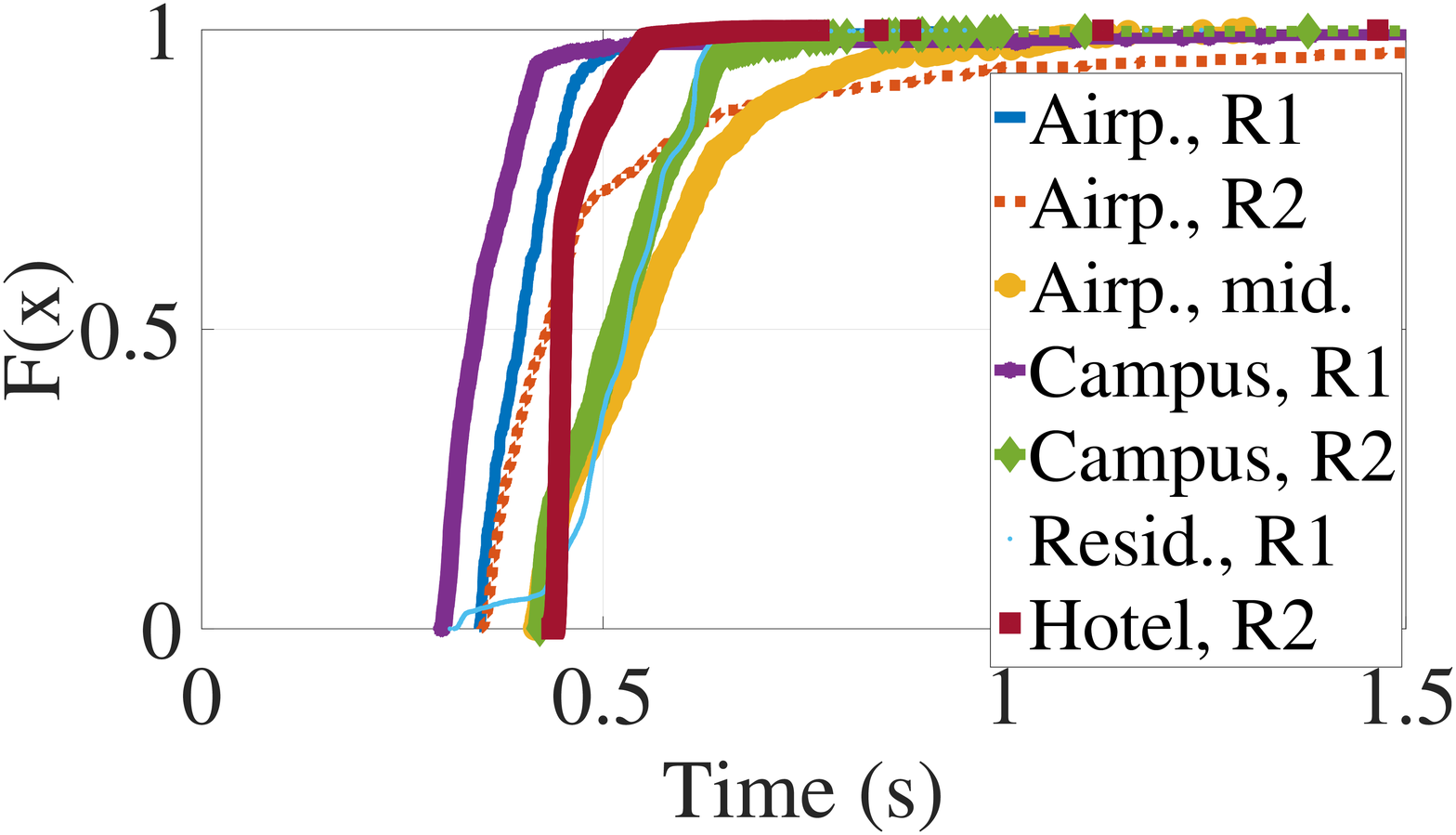}  \label{fig:CDfSeattleNonFlights}
}
\subfigure[3 different flights between Reg.~1 and Reg.~2.]{
\includegraphics[width = 0.45\linewidth]{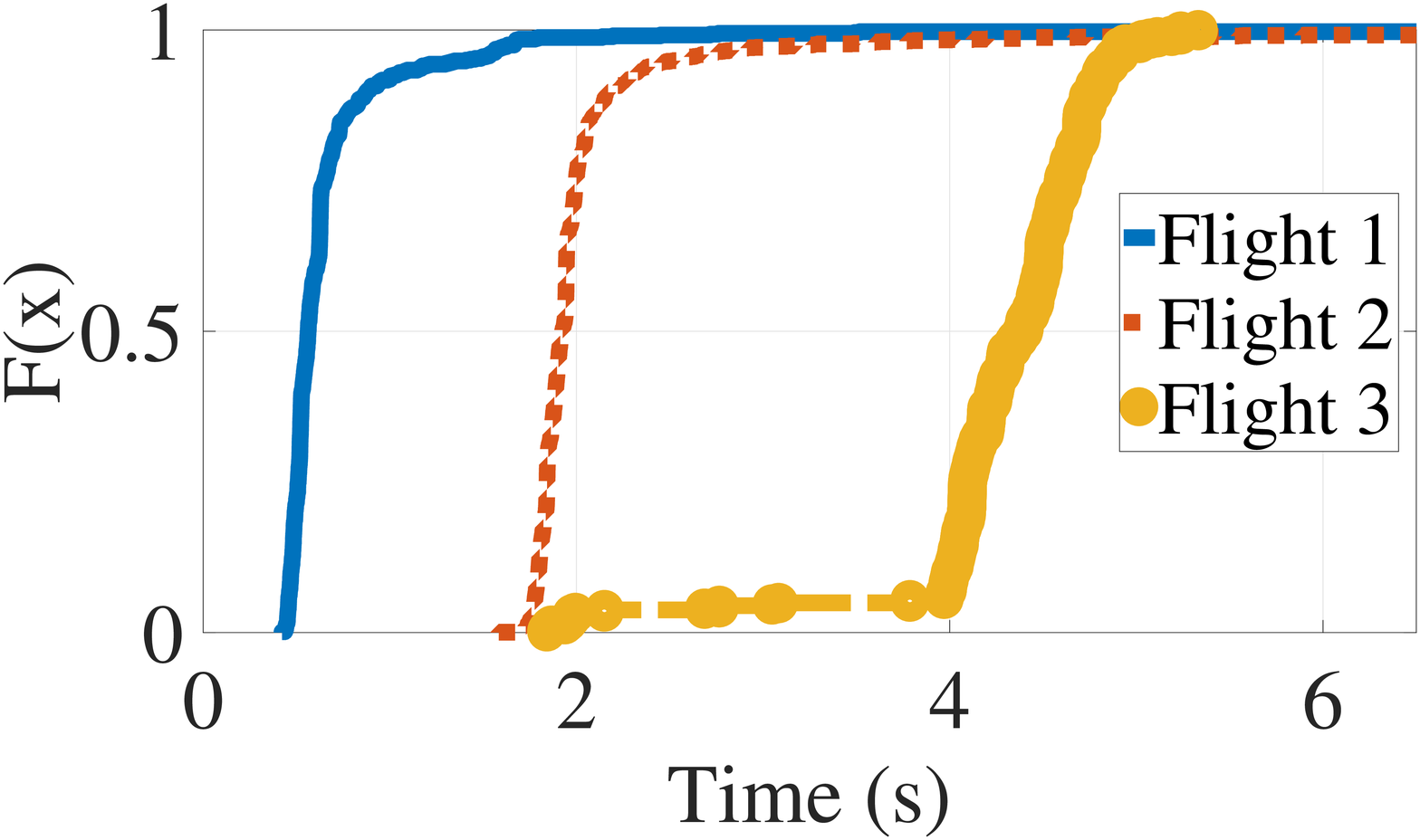} \label{fig:CDFSeattleFlights}
}
\vspace{-0.25cm}
\caption{Task completion latency CDFs, for the same execution option requested from 10 different locations in Region~1, Region~2, and en route in-between the regions.\label{fig:CDFsSeattleTrip}}
\vspace{-0.3cm}
\end{figure} 

To understand the variability of task completion latencies across different edge conditions and locations, 
we measured the latencies experienced by an edge node carried by a human on a trip from Region~1 to Region~2. 
The experiment included 16.3 hours of measurements collected in 10 different environments; the edge node was accessing the benchmarks over WiFi and cloud connectivity available in the given environment. 

Table~\ref{table:TripToSeattle} lists the environments we captured: 
a campus, a residential location, and an airport in Region~1; 
two flight segments and a stop-over airport in the middle between Region~1 and Region~2; 
a corporate campus, a hotel, and an airport in Region~2, 
and one segment of a flight between Region~2 and Region~1. 
Task completion latency CDFs captured in these environments for the PIC task with option 3 executed in an AWS instance in Region~1 (option 3 in Table~\ref{table:DevicesInExperiments}), are plotted in Fig.~\ref{fig:CDFsSeattleTrip}. 
For the same task executed in both Region~1 and Region~2, Table~\ref{table:TripToSeattle} summarizes the key properties of the task completion latencies observed from the different locations -- the median ($m$), the $10^{\textrm{th}}$ and the $90^{\textrm{th}}$ percentiles ($p_{10}$ and $p_{90}$), the span between the $10^{\textrm{th}}$  and the $90^{\textrm{th}}$ percentiles ($sp$), all in units of seconds, and the number of samples ($N$). 

Comparing task completion latencies in Regions 1 and 2 as experienced from the different locations, 
we observe, as we expected, that in most cases the median task completion latency is smaller when the task is executed in a nearby datacenter -- this is the case for locations 1, 3, 7, 8, and 9, for example. 
We note, however, that \emph{while task completion latency is related to the geographical proximity of an execution point, geography does not fully determine the latency}. 
For example, for the three Region 2 environments we examine (locations 7-9 in Table~\ref{table:TripToSeattle}), the inter-location difference for the Region~2 datacenter execution is comparable to the difference between Region~1 and Region~2 executions. 
Also, for one of the locations we examine, location~2, the median task completion latency in its regional datacenter is slightly higher than in the remote one; in this case, as the spans for Region~1 and Region~2 are vastly different as well ($sp$=0.17 vs.~$sp$=0.43), phenomena other than geographic distances play important roles in determining task completion latency. The difference in the $sp$ values is also pronounced for location 5 ($sp$=0.43 vs.~$sp$=0.99). This suggests that both local conditions and wider networking connectivity and execution point conditions are important for determining task completion latency as well.     

We observe that the extent of the concentration of task completion latency around the median varies widely in different conditions. 
This can be observed from the difference in the shape of the time completion latency CDFs shown in Fig.~\ref{fig:CDFsSeattleTrip}, and from the range of $10^{\textrm{th}}$ to $90^{\textrm{th}}$ percentile span $sp$ values shown in Table~\ref{table:TripToSeattle}. 
We also note that the distributions are tight, with $sp\leq 0.11$, on one of the campuses we examine, and, somewhat surprisingly, in one of the airports and in the Region 2's hotel; the lack of jitter in the time completion latencies suggests the presence of well-provisioned reliable WiFi and location-to-cloud networking one would not normally associate with airport or hotel connectivity. 

Overall, our experiments demonstrate that for latency-sensitive tasks, as both local and local-to-cloud connectivity affect the conditions in non-straightforward ways, predicting the median latency and the CDF concentration is very difficult. Based on these results, we argue that \emph{experimental characterizations are needed to accurately capture time completion latency characteristics in a given environment}. 

Our experience with latencies of in-flight task completions, as 
depicted in Fig.~\ref{fig:CDFsSeattleTrip}(b), showcased the potential for task completion latencies to vary dramatically in challenged environments -- \emph{the median task completion latencies for the in-flight environments differ by 11x}.
Such wide range of latencies suggests that for in-flight and other challenged environments, little can be said about task completion latencies without an empirical characterization.\footnote{Coincidentally, our experience also showcased the difference between ``basic'' and ``high-speed'' in-flight WiFi. 
On one of our flights, the median task completion latency was only 160~ms higher than the median latency recorded in the flights' departure airport (a 40\% increase). 
On the other two flights, the median latencies were $1.5$~s and $3.9$~s higher than the median latencies in the departure airports (\emph{$>$300\% and $>$700\% increases}, correspondingly). 
We deduced that one of the flights happened to be among the 7.2\% seat-miles of flights that offer high-quality WiFi~\cite{RouteHappy2018}. 
}  

\section{Task Completion Latency Stability\label{Sect:ResponseTimeStability}}
\par 
In our experiments, we observed, with some surprize, 
that  the statistics of task completion latencies in a given environment remained largely the same for our multi-hour experiments. While wireless channel variations introduced disruptions, the underlying fog system response times appeared to be somewhat stable.  

To examine this further, we conducted a 10-day experiment with the 
FSP benchmark executing in an AWS EC2 instance in a Region 2 datacenter ($>$800,000 consecutive executions), 
and a 2-day experiment with the PIC benchmark executing in an AWS EC2 instance in a Region 1 datacenter ($>200,000$ consecutive executions). To exclude WiFi-related disruptions, we conducted these experiments with an edge node connected via Ethernet. 

\begin{figure}[t]
  \centering
  \subfigure[FSP, lower network load.]{
  \includegraphics[width=0.45\linewidth]{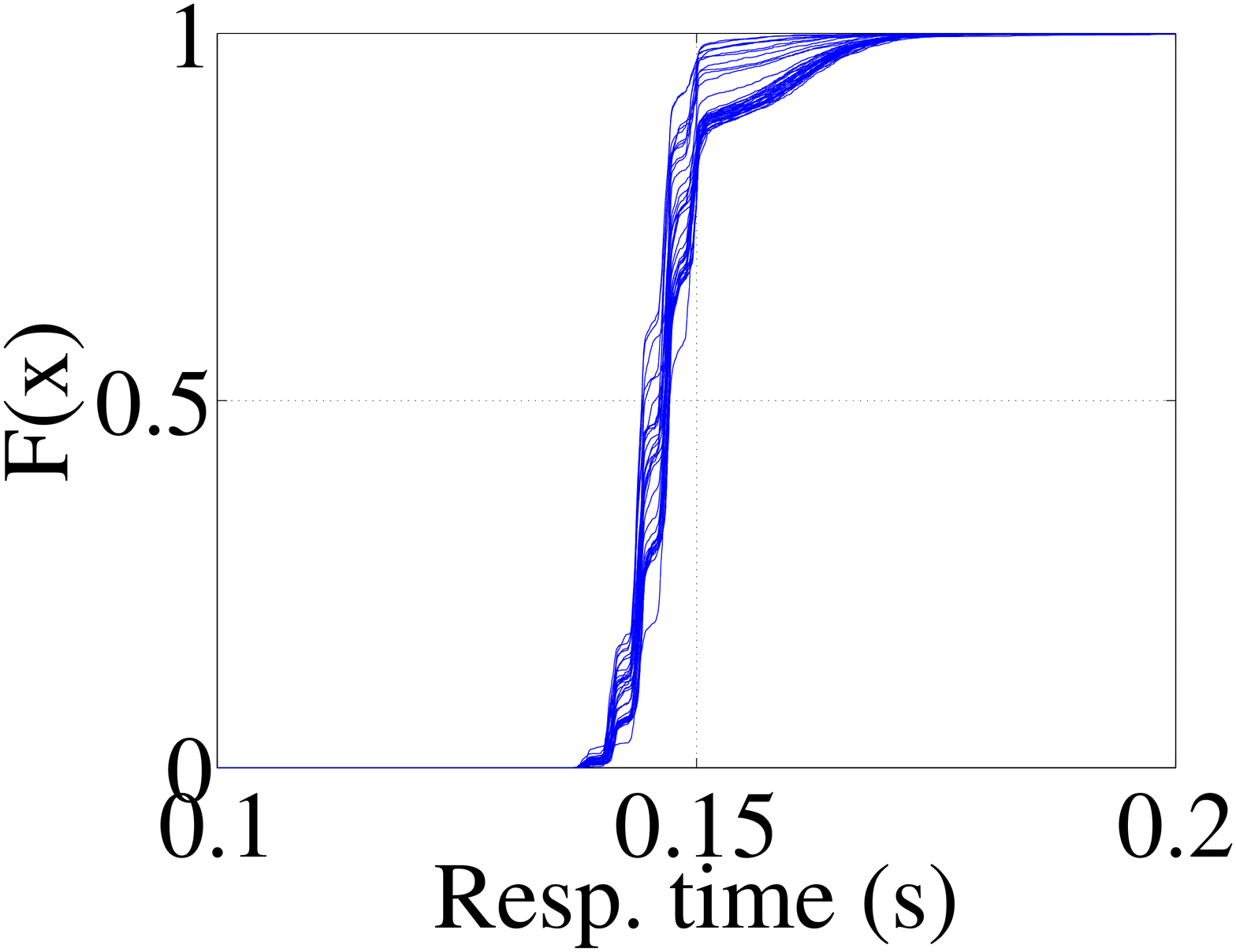}\label{fig:CDFStable}}
  \subfigure[FSP, higher network load.]{
  \includegraphics[width=0.45\linewidth]{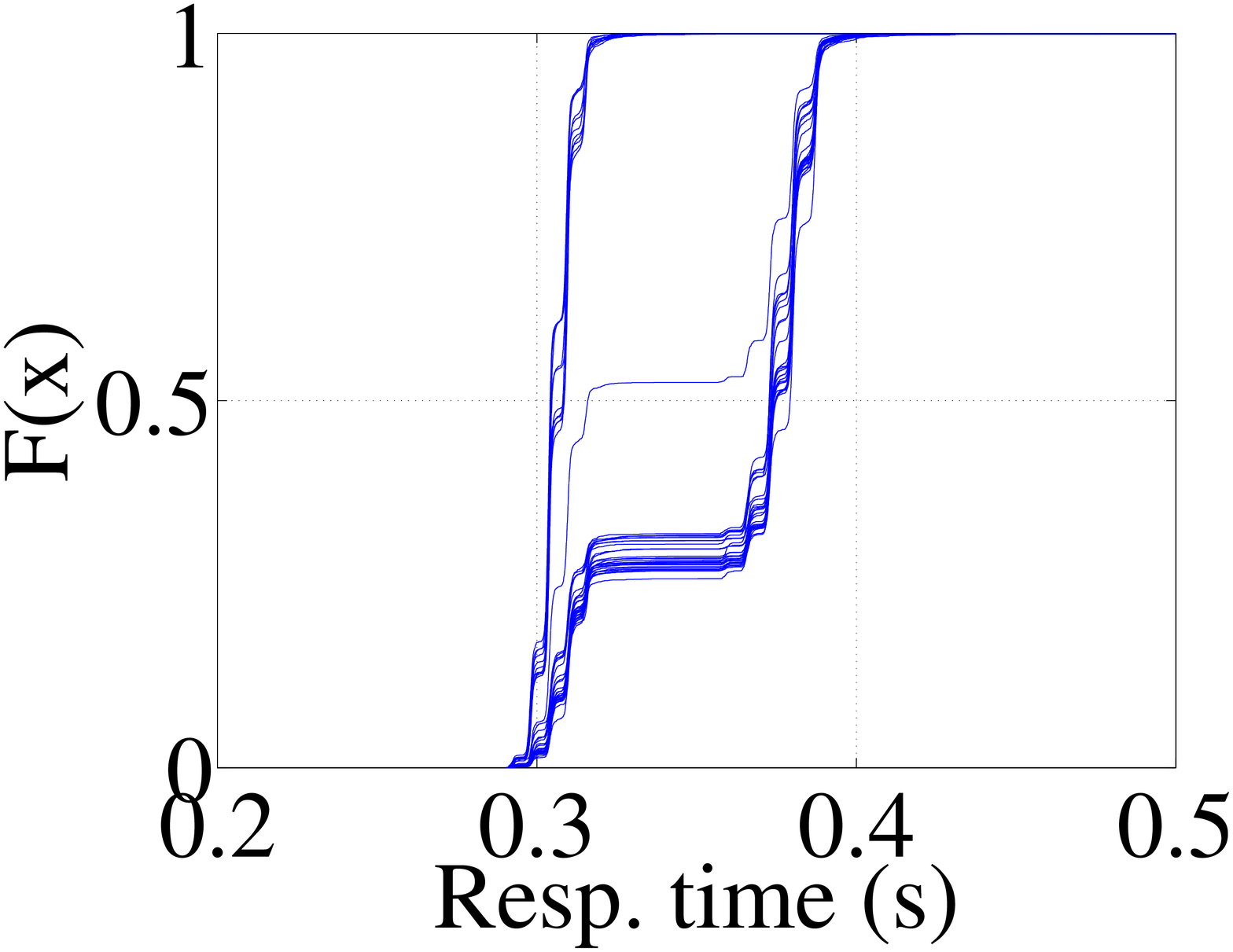}
  \label{fig:CDFBifurcated}}
  \vspace{-0.3cm}
  \caption{The overlayed CDFs of 5,000 consecutive task executions intervals in a 10-day experiment, for an AWS EC2 instance-based execution point, for the FSP benchmark with lower (a) and higher (b) network utilization. Over the 10-day experiment, latency characteristics of the execution point remained stable for one experiment (a), and changed once for the other (b).}
\end{figure}

\par
Our results demonstrate that task completion latencies' properties remain stable for long periods of time, undergoing changes only once in a while. 

For example, for the FSP benchmark executed with the option of a small dataset transmitted over the network (500 lines transmitted), the statistical properties of task completion latencies \emph{remained stable over the entire 10-day interval}. 
Fig.~\ref{fig:CDFStable} shows, for this case, the CDFs for the successive 5,000-invocation intervals of this experiment, overlayed with each other. 
Throughout the 10 days of this experiment, the $5^{\textrm{th}}$--$75^{\textrm{th}}$ percentiles of task completion latencies stayed within 2.2\% from each other, the $90^{\textrm{th}}$ and the $95^{\textrm{th}}$ percentiles -- within 8.5\%. 
For the FSP mini-benchmark executed with the option of a larger dataset transmitted over the network (10,000 lines transmitted), \emph{over the 10-day course of the experiment, the CDF statistics changed once}, on the $4^{\textrm{th}}$ day of the experiment. Fig.~\ref{fig:CDFBifurcated} shows, for this case, the CDFs for the successive 5,000-invocation intervals of this benchmark. 
The difference between the percentile values for the different CDFs in this case is up to 20\%; 
for the CDFs before and after the $4^{\textrm{th}}$-day transition the difference in all percentiles does not exceed 2.1\%.

\par 
For the PIC mini-benchmark, the CDF statistics changed only once as well, in a 2-day interval, 
but the changes were more pronounced: for one of the options, the median task completion latency increased 6x (from 0.05~s to 0.32~s), and for the other -- 10x (from 0.29~s to 2.9~s). We hypothesize that this notable change is related to CPU sharing in the AWS t2.micro instance we use, as the PIC is a CPU-sensitive benchmark. Similarly to the case of the FSP, before and after the transition, task completion latency CDFs remained relatively stable. Over the 17 hours before the transition and the 27 hours after the transition, the CDFs' quantiles differ by no more than a few percent.

\par 
The observed stability in task completion latencies indicates that the execution point latency characterizations can be conducted relatively infrequently, but do need to be updated periodically. In particular, the extent of task completion latency changes we observed for the PIC task -- e.g, from 0.29~s to 2.9~s -- is \emph{on the order that would call for selecting a different task execution point for a latency-sensitive task}. Our observations suggest that in fog systems with shared resources, latency-sensitive tasks may need to be re-assigned as system conditions change.

\section{Serverless Execution \label{sect:ServerlessExecutionNotes}}

We studied the properties of serverless execution with both AWS Lambda and Microsoft Azure serverless functions, as described in Section~\ref{Sect:OurExperimentalSetup}. Below we first comment on serverless computing support of differing numbers of users, then describe the associated task execution latency characteristics unique to serverless computing. 

\subsection{Serverless computing options as infinite-capacity execution points}

One of the core advantages of serverless computing is \emph{auto-scaling}: unlike traditional cloud computing resources, which need to be created ahead of time, serverless instances are spun up and down automatically based on user demand. This saves application developers and administrators from worrying about creating the right number of processes for the users. 
From the point of view of task assignments in fog computing architectures, this property of serverless execution translates to having \emph{execution points with infinite capacities}. 

We appreciated the auto-scaling property of serverless computing when conducting our benchmarking experiments: we had to keep track of whether we instantiated the conventional services, 
but did not have to worry about serverless ones. 
The downsides of enabling auto-scaling in serverless computing are some limitations in languages and functionality, the need to adapt programs to serverless APIs (which we had to do for our benchmarking experiments), and complex latency characteristics of serverless execution, as we describe below.

\subsection{Latency changes with autoscaling and spin-down}

\par 
Due to cloud providers internal resource allocation mechanisms that prioritize in-demand functions over functions accessed infrequently, serverless functions that are called more often execute faster. 
Practitioners who work with serverless computing have been noting this~\cite{KeepingLambdasWarm2017}; 
however, we are unaware of attempts to incorporate properties of serverless execution points with fog resource allocation mechanisms. 
From the point of view of assigning tasks to execution points, serverless function autoscaling leads to a paradoxical, counter-intuitive behavior: \emph{the more work we assign to a serverless execution option, the faster it responds}. 


The differences between task completion times for functions invoked with different frequencies are substantial. 
For example, in our experiments with the PSF task in the Region 1 AWS datacenter, the difference between the median execution times of functions invoked within 1.5~s of each other and function invoked with longer delays ranged from 7\% to 42\%. 

\begin{figure}[t]
  \centering
  \subfigure[Task completion latency vs. $\Delta t$.]{
  \includegraphics[width=0.45\linewidth]{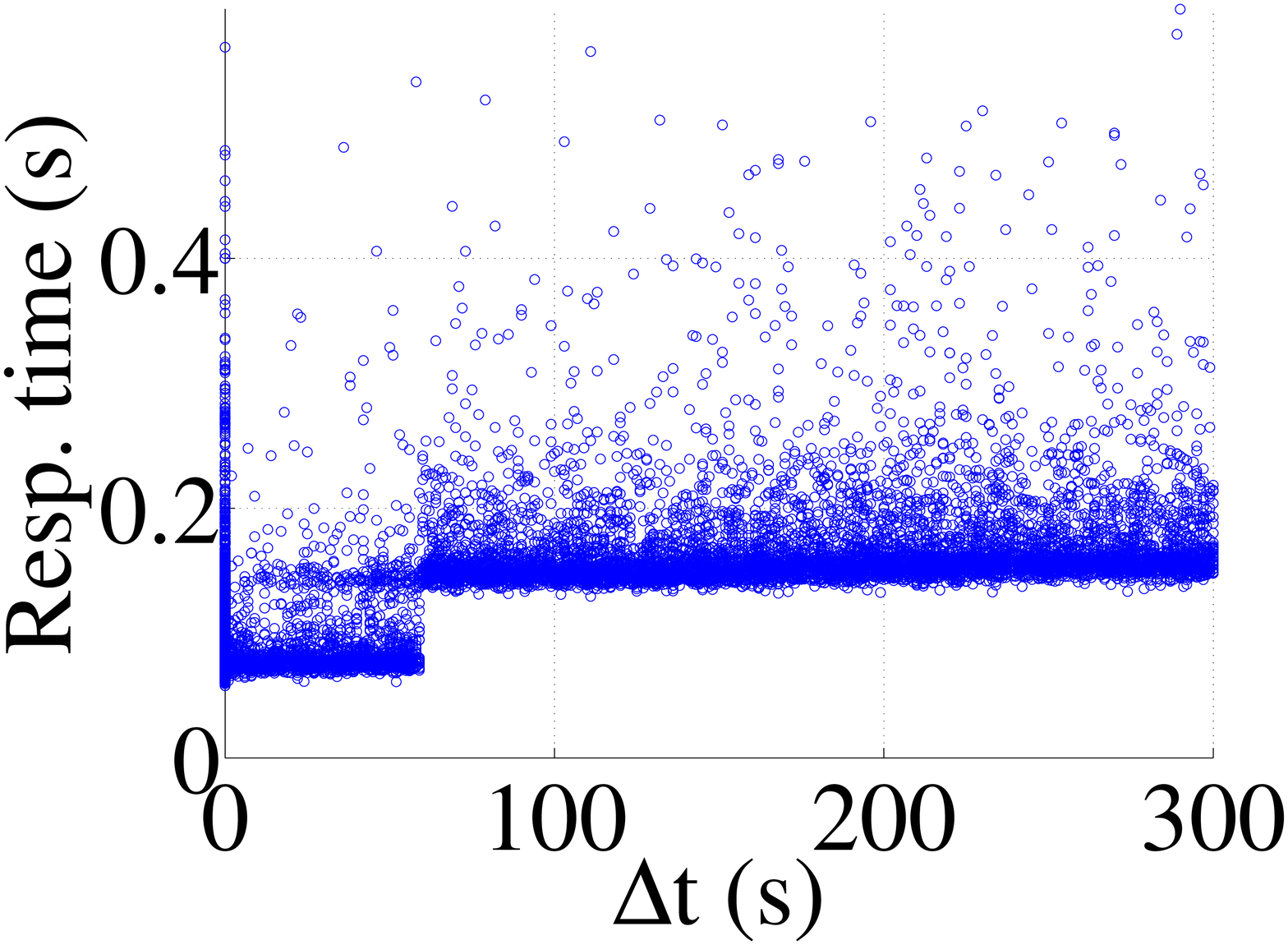}\label{fig:Autoscaling_Azure}}
  \subfigure[CDFs for the three cases.]{
  \includegraphics[width=0.45\linewidth]{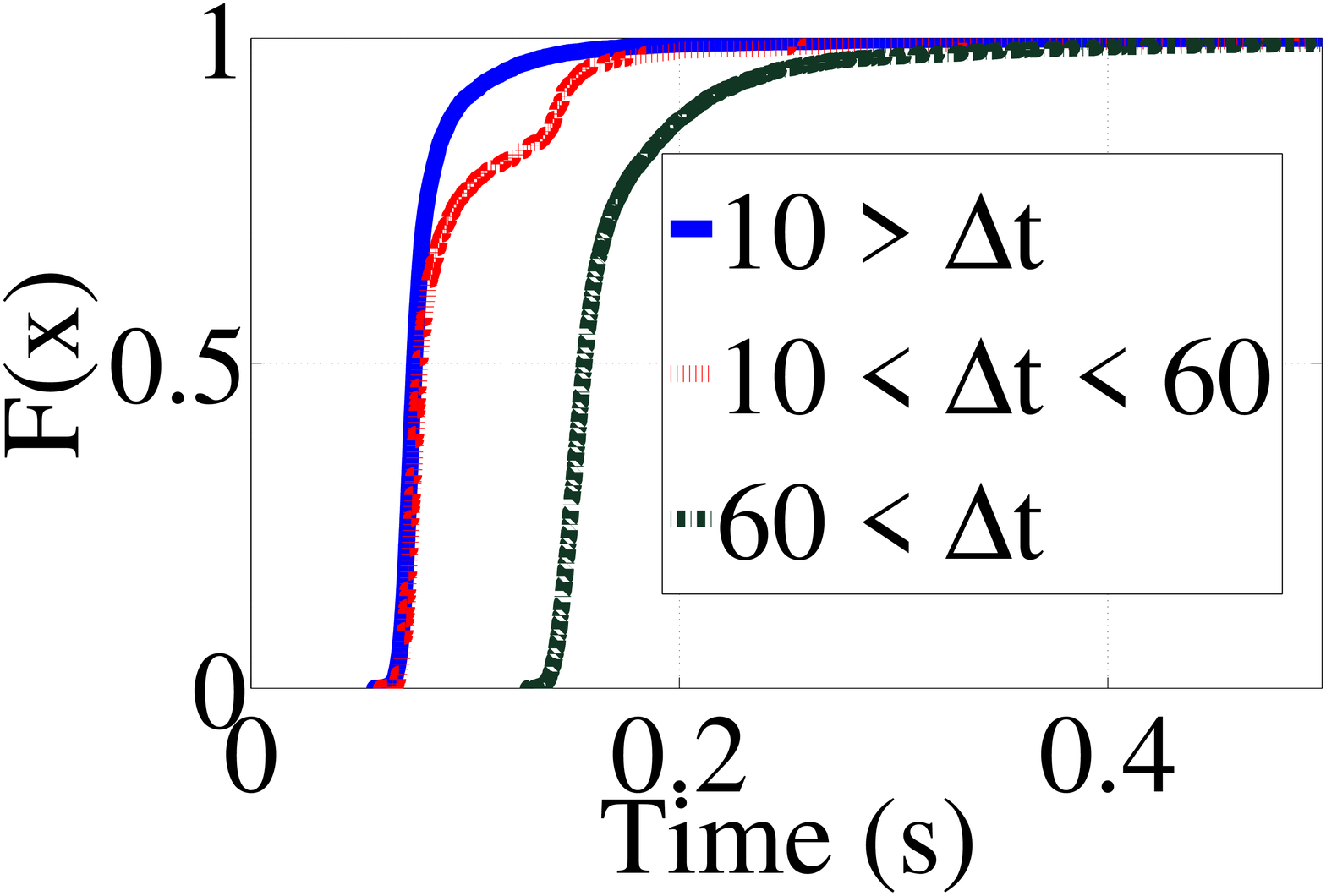}
  \label{fig:Autoscaling_AzureCDFs}}
 \subfigure[PDFs for the three cases.]{
  \includegraphics[width=0.45\linewidth]{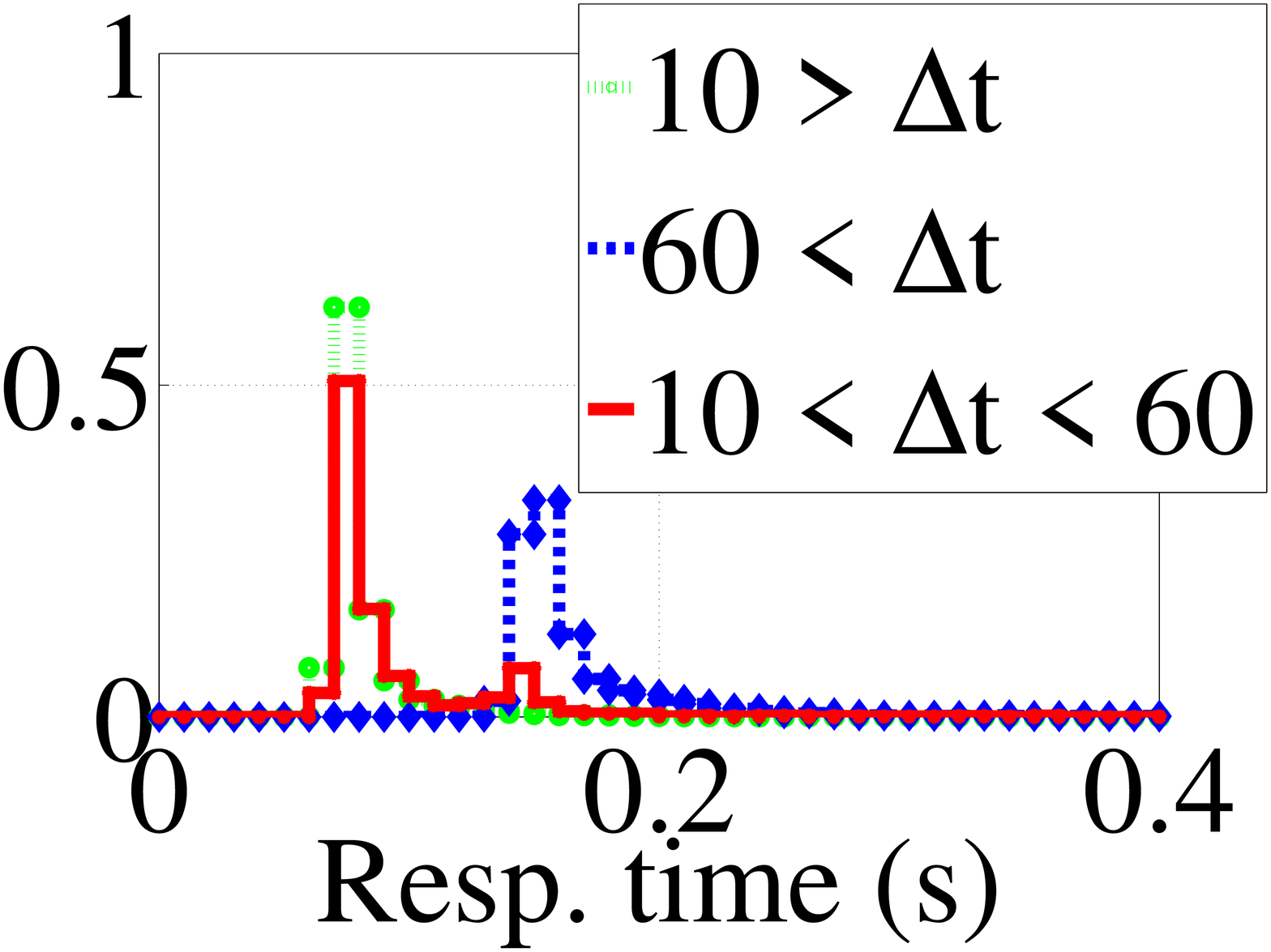}
  \label{fig:Autoscaling_AzurePDFs}}
  \vspace{-0.3cm}
  \caption{Task completion latencies for a Microsoft Azure serverless function, for the different times between function invocations $\Delta t$: (a) Scatterplot of task completion latencies vs.~$\Delta t$, and the CDFs (b) and the PDFs (c) for 3 distinct response time dynamics we identified in this data. 
  }
\end{figure} 



When serverless functions are not invoked for longer periods of time, 
task completion latencies can increase further. For example, in our experiment with invoking an AWS function in a Region 1 datacenter every 30 minutes over the course of a day, the median response time was 
0.95~s, for a function that responds in 0.5~s when called frequently -- a 90\% increase in latency. In our experiment with invoking a Microsoft Azure function in the Region 1 datacenter every 30 minutes over the course of 1.5 days, the average response time was 6.0~s, for a function that normally responds in under 0.15~s -- a \emph{40x difference}. The differences we observe between these ``cold start'' times with different serverless platforms have also been recently noted in~\cite{McGrath2017}. The wide variety of latency values possible with serverless computing in different circumstances 
suggests that serverless execution point latencies need to be empirically obtained. 

\subsection{Modeling latency in the presence of spin-down}

\par
To understand the potential approaches to modeling spin-down latency increases, we conducted a longer-term experiment with one particular execution option: 20,000 consecutive Microsoft Azure function invocations in Region 1 with randomly chosen time between the invocations $\Delta t$, with the maximum $\Delta t$ = 5~min (300~s). This experiment took approximately 30 days. The experiment focused on the effects of spin-down, rather than on service scale-out; we invoked the serverless function from only one location, thus the time between invocation was never less than the minimum function response time of 0.06~s. 
The scatterplot of the obtained response times versus $\Delta t$ is shown in
Fig.~\ref{fig:Autoscaling_Azure}.\footnote{This plot excludes 64 outlier points with response times $>0.5$~s.}

In this data, we were able to isolate three distinct cases: 
$\Delta t<$~10~s ($N=10,425$),
10 s $< \Delta t <$~60~s ($N=1,641$),
and $\Delta t >$~60~s ($N=7,877$). The CDFs and the PDFs for these three cases are shown in Fig.~\ref{fig:Autoscaling_AzureCDFs} and Fig.~\ref{fig:Autoscaling_AzurePDFs}. 
The task completion latencies for the cases of $\Delta t<$~10~s and $\Delta t>$~60~s are drawn from different distributions; the task completion distribution for the case of 10 s $< \Delta t < $ 60 s has the statistical properties of the mixture of the other two distributions, as can be observed, for example, in Fig.~\ref{fig:Autoscaling_AzurePDFs}. 

Our experiments suggest that the underlying mechanisms controlling task completion latencies in serverless execution are complex. It is clear that modeling these latencies requires knowledge of function inter-invocation times; however, individual edge nodes may not have the information about the frequency of invocation requests generated by other nodes, in which case they can only provide pessimistic latency estimates. Moreover, to empirically characterize the latency associated with a specific serverless execution point, the point \emph{needs to be probed at the expected task invocation frequency}, as the observations made at other frequencies may be substantially different.

\section{Co-optimizing task quality and timeliness\label{Sect:Model}} 

In this section we present our framework for task quality-latency co-optimization, 
compare and contrast two elements of the framework with the traditional techniques, 
and demonstrate that the latency characterizations this framework requires can be obtained relatively easily. 

\subsection{Model}

\par
We co-optimize task completion latency and quality by \emph{optimizing task utilities that have both an intrinsic quality and a time-dependent component} -- the approach that is inspired by the approaches previously developed in the field of anytime algorithms~\cite{zilberstein1996using}. In selecting execution points for each task $j$, our goal is to maximize the expected utility 
\begin{equation} 
U_{jx} \buildrel\triangle\over = \mathbb{E}[ Q_{jx}(T_{x})], 
\end{equation} 
where $Q_{jx}(t)$ is the \emph{utility obtained when the execution option $x$ for task $j$ completes in time $t$}, and $U_{jx}$ is the expectation of $Q_{jx}$ over $T_x$, the random variable representing the completion time of execution option $x$. 

We define $Q_{jx}(t)$ to be a combinations of an \emph{intrinsic utility} $A_{jx} \in [0,1]$ and \emph{time-dependent utility} $f_j(t) \in [0,1]$. The time-dependent utility functions $f_j(t)$ can be different for different tasks. Multiple different time-dependent utility functions are showcased in each of~\cite{zilberstein1996using,RealTimeSystemsBook,Sun2017Update}, for example; three notable important functions categories of them are the following: 
    
    \textbf{1. Step function}: $f_j(t) = 1$ for $t\leq t_v$, $0$ otherwise. 
    This function can be used when a task needs to be completed under a deadline $t_v$ (i.e., a hard deadline for task completion), but no additional utility can be gained from completing the task sooner. 
    
    \textbf{2. Decaying function}: 
    A monotonically decaying $f_j(t)$, such as $f_j(t) = e^{-t}$, corresponds to a preference for a faster system response without a firm execution deadline~\cite{Sun2017Update}.  
    
    \textbf{3. Wait-readily-first function}: 
    In many realistic scenarios, the time-dependent utility function $f_j(t)$ is constant for some $t \leq t_v$, and is decaying for $t > t_v$.
    In control systems and other machine-to-machine interactions, this is the case where the system is rate-limited by its other components, e.g., its sampling rate or its response dynamics. 
    This function also applies to human-facing systems, as humans perceive delays below certain values as instantaneous, and see 
    systems as progressively losing quality 
    as delays increase~\cite{miller1968response}. 
    In our numerical results, we use a piecewise linear function to model this class of functions, specifically $f_j(t) = 1$ for $t\leq T_e$, $1-\frac{t-T_e}{T_s-T_e}$ for $ T_e\leq t \leq T_s$, and $0$ for $ T_s \leq t$. 

\par
Time-dependent utility functions can be different for the same task. 
In classic control, for example, different system dynamics and stability parameters call for different response timelines~\cite{ControlAsAServiceReportInaltekin2017}. 
In modern video processing applications, as noted in~\cite{ran2018deepdecision}, as the rate of difference between different frames in a video feed increases or decreases, video processing timeline requirements change. 
    
\par
Time-dependent utility functions can also be different for different human users. Generic commonly cited rule-of-thumb characterizations of human experiences with computer systems' latencies have long been established~\cite{miller1968response}; 
however, multiple ongoing studies have been demonstrating how multiple factors affect delay perception and tolerance~\cite{galletta2004web,Caldwell2009Delays} -- for instance,~\cite{Caldwell2009Delays} found that the delay tolerance is affected by task complexity, importance, situation urgency, and time availability.\footnote{Scientists even examined, but were unable to conclusively establish, the dependency of delay tolerance on gender and personality type~\cite{Wang2002User}.} 
Long-term, we can envision automatically collecting individual user feedback on system latency (similar to the service quality surveys currently done on groups of users, e.g., in~\cite{Chen2017Empirical}) to understand personalized latency-related preferences of the individual users.

\subsection{Task completion time as a random variable}

While experimental characterizations of fog computing systems often include latency CDFs~\cite{Chen2017Empirical,Perez2017Experimental}, studies that examine task assignments usually do not take execution time randomness into account. 
The traditional approach is to treat the execution option as characterized by one statistic. 
Representing a random variable by one of its statistics is a common simplification, used across multiple engineering disciplines; however, in the context of modern fog computing systems, it suffers from two major drawbacks: 

\textbf{1. Implicit assumption of CDF uniformity}: 
Operating over one statistic of a distribution 
implicitly assumes that different distributions are linearly shifted versions of one another. 
\emph{Our results emphasize that different task execution options have 
different shapes} (e.g., different 10-90$^{\textrm{th}}$ percentile spans). 
We expect task latency completion times' CDFs to vary even more in the future, 
as additional heterogeneous nodes become a part of fog computing platforms~\cite{Georgiev2016LEO,Zahran2017Heterogeneous}.  

\textbf{2. Inability to handle differentiated timeliness preferences}: 
When only one statistic of a distribution is used, it is not possible to provide differentiated, appropriately selected services to users with different time utility functions. 

Our framework addressed both concerns by treating execution delays as random variables. 

\subsection{Time-quality co-optimization in comparison to traditional approaches\label{Sect:CoOptimizationTimeQuality}}

One widely used approach in fog computing is to seek to \emph{minimize the delay} associated with task assignment, subject to additional constraints -- this is the objective in~\cite{Souza2016Towards,Jia2017Optimal,Xiao2017QoE,Tan2017Online}, for example. This approach does not take into account the possibility of different quality of results. Moreover, seeking to minimize the delays does not always lead to good solutions for timeliness functions for which latencies under some value $t_v$ are equally valuable, such as the step function and the wait-readily-first function. For these functions, minimizing the delays leads to over-constraining the solutions, and not allowing for trade-offs between the tasks. 

An approach that does allow for quality differentiation is \emph{maximizing the quality}, subject to delay and other constraints; variants of this approach are explored in~\cite{Han2016MCDNN,ran2018deepdecision}. 
This approach implicitly assumes that the latencies of the different execution points have the same characteristics, and does not allow operating over different timeliness preferences of the users. Specifically focusing on maximizing the quality, it can result in selecting options with small quality gains over options with slightly lower quality but more robust timeliness performance.


\subsection{Learning task completion times}

\begin{figure}[t]
  \centering
  \subfigure[A CDF and its approximations.]{
  \includegraphics[width=0.6\linewidth]{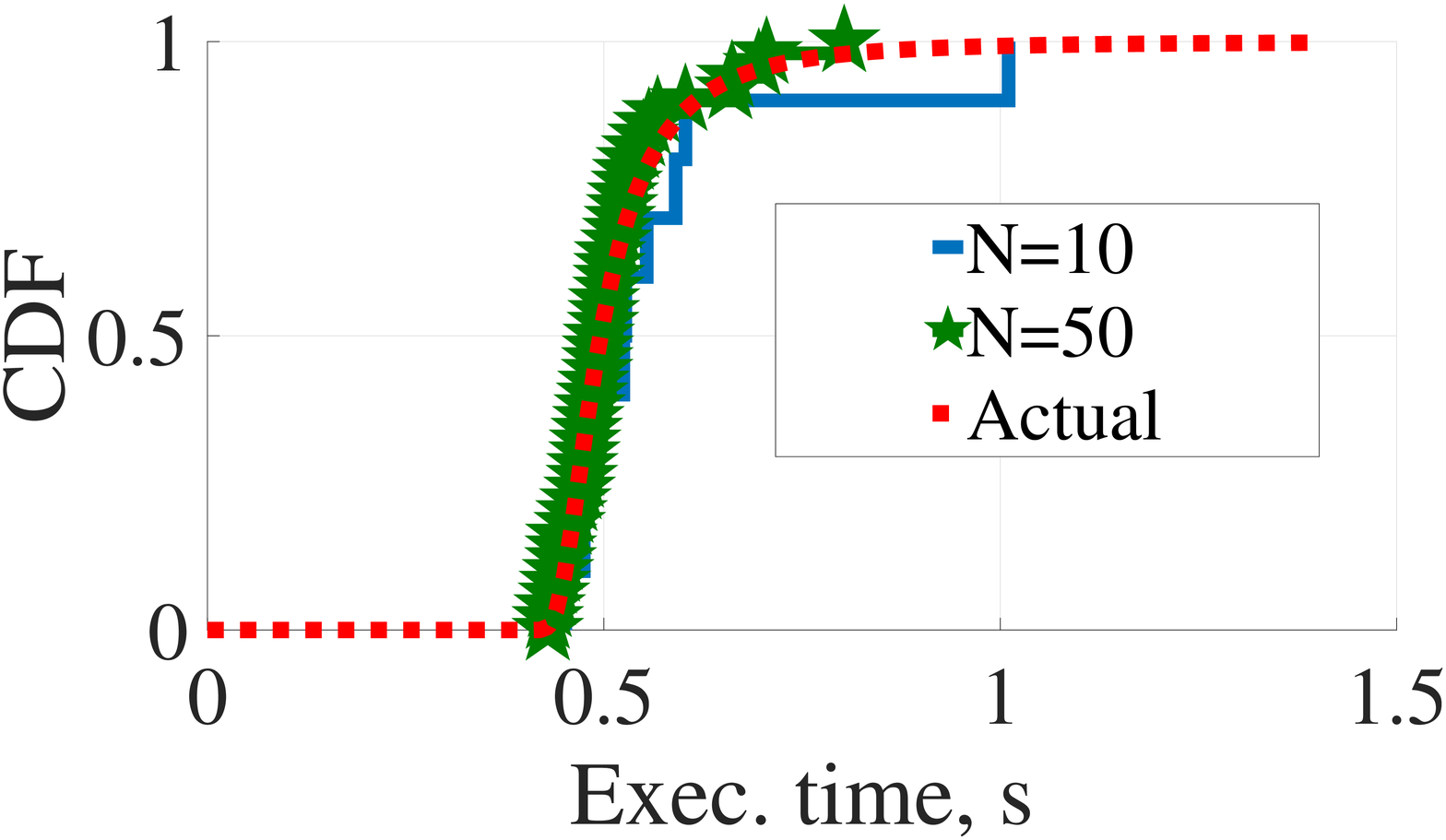}}
  \subfigure[Average error.]{
  \includegraphics[width=0.45\linewidth]{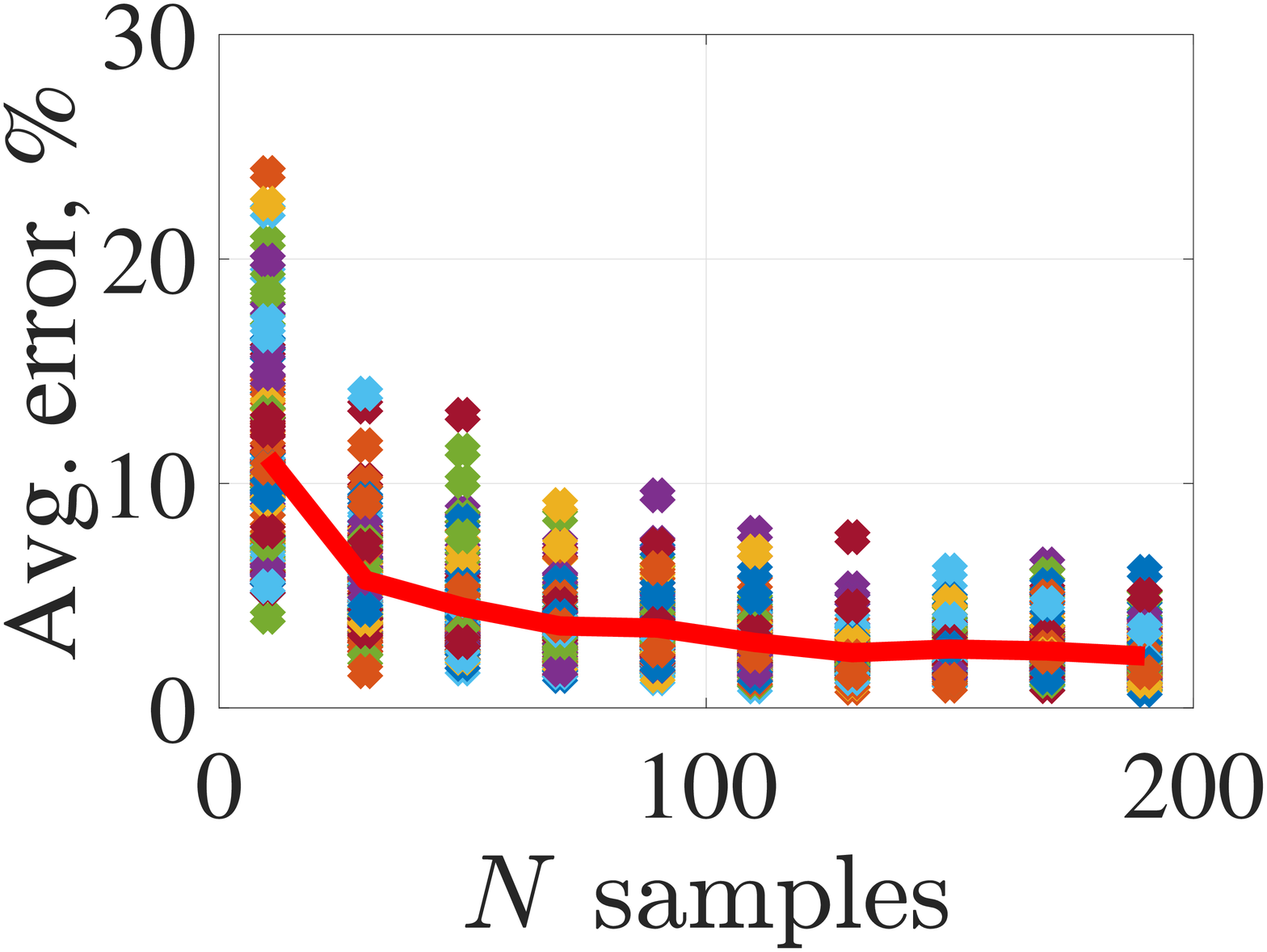}}
  \subfigure[Maximum error.]{
  \includegraphics[width=0.45\linewidth]{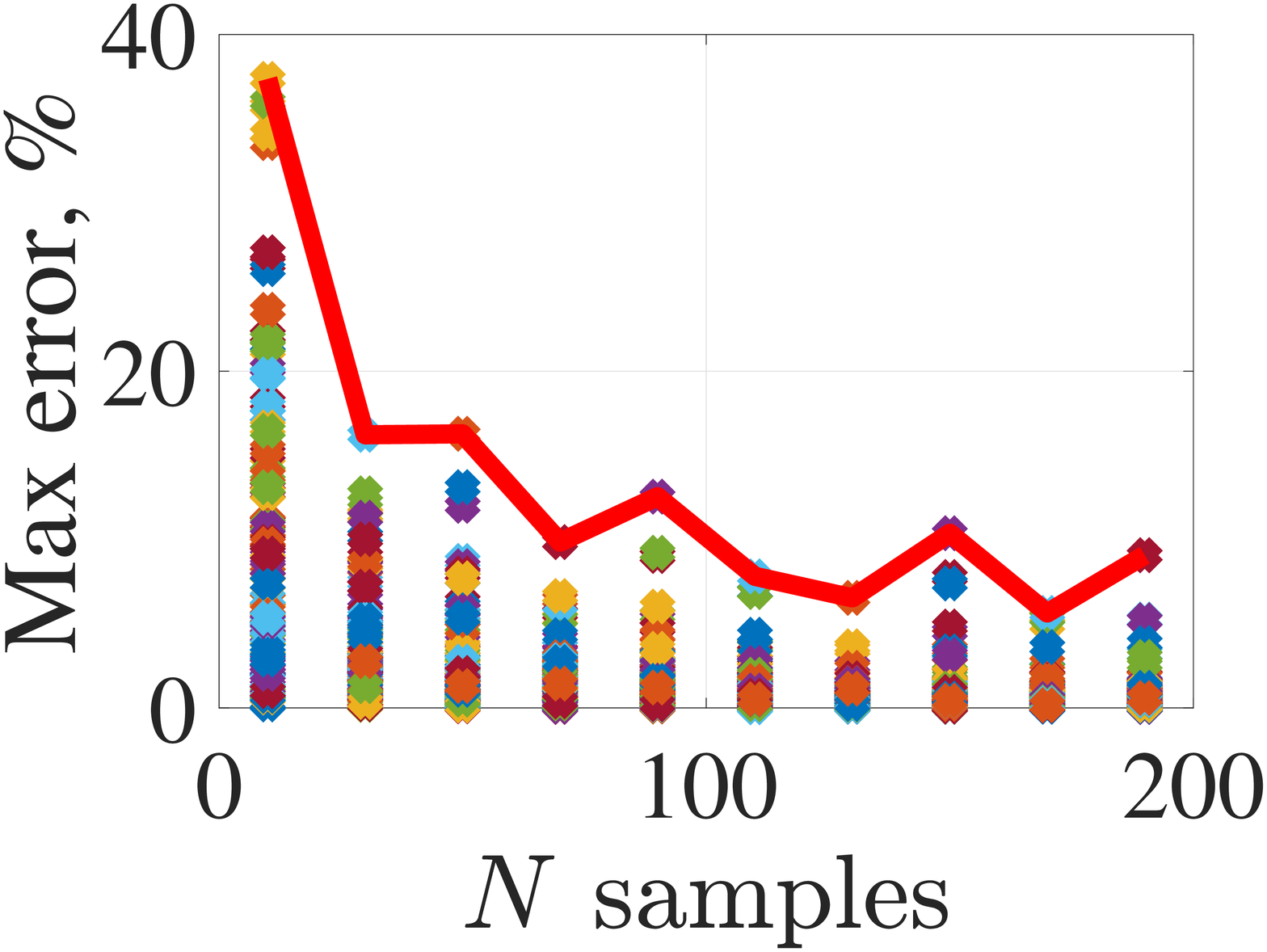}
  }
  \vspace{-0.3cm}
  \caption{An experimentally obtained execution point latency CDF $F$ and its estimates $\hat{F}_N$ based on $N=10$ and $N=50$ samples from the distribution (a), and the average (b) and the maximum (c) distance between the experimental $F$ and its estimates $\hat{F}_N$, for the different number of samples $N$. Even a relatively small $N$ allows characterizing a CDF well. \label{fig:CDFErrorsGEV}}
\end{figure}

In this work we are arguing for the need to experimentally characterize task completion latencies' CDFs.  

Fortunately, relatively few samples are required to obtain an approximation of a CDF. 
For instance, Fig.~\ref{fig:CDFErrorsGEV}(a) shows an experimentally obtained CDF $F$ for setting 1 in Table~\ref{table:QualityFits}, in comparison with Kaplan-Meier estimates of the CDF, $\hat{F}_N$, 
based on 10 and 50 samples $N$ from this distribution. It can be observed that even for small $N$, $F$ and $\hat{F}_N$ are relatively close to each other. 
Fig.~\ref{fig:CDFErrorsGEV}(b,c) show the average and the maximum distances between $F$ and $\hat{F}_N$ for different $N$ values for this distribution. For each value of $N$, samples were randomly drawn 100 times; the dots on the graphs show the individual results, while the line shows the results' average in Fig.~\ref{fig:CDFErrorsGEV}(b), and the results' maximum in Fig.~\ref{fig:CDFErrorsGEV}(c). 
It can be seen that even a relatively small number of samples allows characterizing a CDF relatively well. 

Obtaining samples to characterize execution point latencies should not be onerous in practice, 
as it is straightforward to measure latencies while executing tasks. Approaches that trade off exploitation and exploration in latency characterization can be developed to further reduce the execution option characterization overhead. 

\section{Application to Task Assignments \label{Application:TaskAssignment}}

We present fog computing task assignment problem formulations in Sect.~\ref{Sect:ProblemFormulations}, and the algorithms we developed to solve the formulated problems in Sect.~\ref{Sect:SolvingUncapacitatedProblem} and \ref{Sect:SolvingCapacitatedProblem}. The numerical results are presented in Sect.~\ref{Sect:SimulationResults}.

\subsection{Problem formulations \label{Sect:ProblemFormulations}}

\begin{figure}[t]
  \centering
   \includegraphics[width=0.7\linewidth]{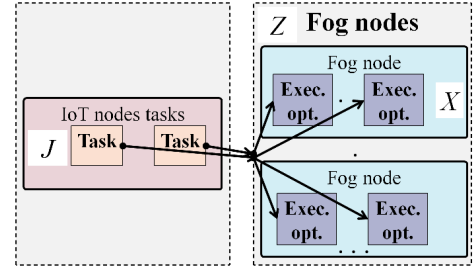}
    \vspace{-0.2cm}
   \caption{ System diagram: tasks $j\in J$ can be serviced by different fog nodes $z \in Z$, each of which can have multiple execution options $x \in X$.
  \label{fig:FogArchitectureSettings}}
  \vspace{-0.2cm}
\end{figure} 

\begin{table}[t!]
\centering
\caption{Notation. \label{table:symbols_used}}
\vspace{-0.3cm}
\begin{tabular}{|p{0.08\linewidth}|p{0.8\linewidth}|}
\hline \small
$z, Z$ & Fog node, fog node set \tabularnewline
$x, X$ & Execution option, the set of execution options \tabularnewline
$j, J$ & Task, task set \tabularnewline
$C_z$ & Service capacity of fog node $z$ \tabularnewline
$I_{jzx}$ & Indicator variable for assignment of task $j$ to fog node $z$ and execution option $x$\tabularnewline
$T_{jzx}$ & End-to-end completion time for executing task $j$ on fog node $z$ with execution option $x$ (seconds) \tabularnewline
$Q_{jx}(t)$ & Utility of executing option $x$ in time $t$ for task $j$ \tabularnewline 
$A_{jx}$ & Intrinsic utility of executing option $x$ for task $j$\tabularnewline
$U_{jzx}$ & The utility of assigning task $j$ to fog node $z$ and execution option $x$ \tabularnewline
$f_j(t)$ & Time-dependent utility for task $j$ finishing in time $t$
\tabularnewline
\hline
\end{tabular}
\end{table}

\par 
We consider the problem of an IoT gateway deciding which points in a fog computing system should process its tasks $j \in J$ -- the settings that are shown in Fig.~\ref{fig:FogArchitectureSettings}. The notation is summarized in Table~\ref{table:symbols_used}. 

\par
In a fog computing system the gateway has access to, the tasks can potentially be served by multiple possible execution points $z \in Z$, that can execute tasks with different execution options $x \in X$. We use $|\star|$ to denote the cardinality of set~$\star$. 
We use indicator variables $I_{jzx} \in \{0,1\}$ to denote the assignment of an execution option $x$ on node $z$ to serve task $j$. We use $\{z^*, x^*\}$ to denote the optimal assignment point and option for a task. We denote the end-to-end completion time of a task $j$ by $T_{jzx}$. We treat $T_{jzx}$ as a random variable with relatively stable characteristics, as substantiated by our experiments, and as elaborated in Section~\ref{Sect:Model}. 
We assume that each fog node $z$ can serve at most $C_z \geq 1$ tasks. This restriction may be related to the number of processes $z$ can run, or to its disk space limitations. For some nodes, $C_z$ can be infinite (e.g., in serverless execution). 

\par
To assign tasks to the best placement points and execution options, we formulate the following optimization problem.

\par \noindent
\textbf{Capacitated Task Admission and Placement (C-TAP) Problem}:
\begin{equation} \label{Pr:CapacitatedPlacement}
\max\underset{j \in J}{\sum} U_{jzx}(T_{jzx})\cdot I_{jzx}
\end{equation}

\begin{align}
\label{c:assignedD} \textrm{s.t.:} \  \ \ \ \  \ \ \sum_{z \in Z}\sum_{x \in X} I_{jzx}  & \leq 1, \  \forall\ j \in J \ \ \ \ \ \ \ \\ \ \ \ \ \
\label{c:capacitated} \sum_{j \in J} \sum_{x \in X} I_{jzx} &  \leq C_z, \ \forall\ z \in Z \\
\label{c:PenaltyMeasure} Pr(f_j(T_{jzx})<q_j) \cdot I_{jzx} & \leq P_j' \\
\label{c:BinaryD} I_{jzx} \in \{0,1\},\ \forall & \ j \in J, z \in Z, x \in X.
\end{align}  
In \eqref{Pr:CapacitatedPlacement}, with a slight abuse of notation, we represented $U_{jzx}$ as a function of $T_{jzx}$ to indicate the dependence of utilities on the distributions of task completion times. Constr.~\eqref{c:assignedD} indicates that each task $j$ needs to be assigned to no more than one fog node $z$ 
and execution option $x$. 
Constr.~\eqref{c:capacitated} indicates that fog node service capacity $C_z$ cannot be exceeded on any node $z$. 
Constr.~\eqref{c:PenaltyMeasure} ensures that the risk of a task not executing in a timely fashion is low: 
it ensures that the probability of a task executing with the utility of time value below $q_j$ is bounded by $P'_j$. 
This constraint restricts the  assignment of some tasks to a subset of execution points. 
Constr.~\eqref{c:BinaryD} defines task assignment variables $I_{jzx}$ to be binary. 
Solving an instance of C-TAP can be seen as jointly addressing admission control and task assignment in fog computing (as some tasks may end up unassigned due to node capacity restrictions). 

The C-TAP is a generalization of a knapsack problem, to multiple choices (as different execution options for each task correspond to different profits $U_{jzx}(T_{jzx})$), multiple knapsacks (specifically, $|Z|$ knapsacks with different capacities), and to assignment restrictions (due to constraint~\eqref{c:PenaltyMeasure})\footnote{In fog computing systems, the assignment restrictions could also arise due to privacy-related and other constraints. The algorithms we develop can be used to solve problems with those assignment restrictions as well.}. The problem is NP-hard in this form~\cite{kellerer2004knapsack}. 

We first demonstrate an optimal solution for the case where all node capacities $C_z$ are infinite, which is the case for serverless execution, and is a reasonable approximation for some high-capacity cloud services. We then demonstrate a dynamic programming-based algorithm that can be used to solve C-TAP problems with a relatively small number of capacitated nodes, which is important in practice as it corresponds to having capacitated local nodes and infinite-capacity cloud resources. 

\subsection{Solving uncapacitated problems \label{Sect:SolvingUncapacitatedProblem}}

\begin{algorithm}[t]
\caption{Local Quality Maximizer (LQM)}\label{alg:LQM}
\begin{algorithmic}[1]
\Procedure{LQM$_z$}{$j$, $T_{jzx}$, $A_{jx}$} \Comment{Calculate best $x$ on $z$}
\For{$x = 1; x \leq |X|; x++$}
\If{$Pr(f_j(T_{jxz}) < q_j) \leq P'_j$}
	\State $u_{jzx} \gets \mathbb{E}(A_{jx} \cdot f_j(T_{jzx}))$
\Else
	\State $u_{jzx} \gets 0$
   \EndIf
\EndFor
\State $u^{best}\gets \underset{x}{\max} (u_{jzx})$, $x^{best}\gets \underset{x}{\textrm{argmax}}(u_{jzx})$
\State \textbf{return} $\{ x^{best}, u^{best} \}$
\EndProcedure
\end{algorithmic}
\end{algorithm}

\begin{algorithm}[t]
\caption{Uncapacitated Assignment (UA)}\label{alg:UncapacitatedAssignment}
\begin{algorithmic}[1]
\Procedure{UA}{$J$, $Z$}
\For{$j = 1; j \leq |J|; j++$}
    \State $\{z^{best}_{j}, x^{best}_{j}, u^{best}_{j} \}\leftarrow ITP(j, Z)$
\EndFor
\State \textbf{return} 
$\{z^{best}_j, x^{best}_j, u^{best}_j \}$ \ \ $\forall j$
\EndProcedure
\Procedure{ITP}{$j, Z$} \Comment{Search for the best $z$}
\State $z^{best}\gets \varnothing$, $x^{best}\gets \varnothing$, $u^{best}\gets 0$
\For{$z=1; z \leq |Z|; z++$}
    \State $\{\hat{u}, \hat{x}\}\leftarrow LQM_z(j, T_{jzx}, A_{jx})$ \Comment{Call LQM on $z$}
    \If{$\hat{u} > u^{best}$}
    	\State $z^{best} \gets z$, $x^{best}\gets \hat{x}$, $u^{best}\gets \hat{u}$
    \EndIf
\EndFor
\State \textbf{return} $\{z^{best}, x^{best}, u^{best}\}$
\EndProcedure
\end{algorithmic}
\end{algorithm}

When all fog node service capacities $C_z$ are infinite, constraint~\eqref{c:capacitated} is relaxed, and the optimization problem becomes an \textbf{Uncapacitated Task Admission and Placement (U-TAP) Problem}. Due to the absence of coupling constraints, the U-TAP can be solved optimally subject to constraints \eqref{c:assignedD}, \eqref{c:PenaltyMeasure}, and \eqref{c:BinaryD} by deciding, independently, on the best $\{z^*,x^*\}$ for each $j$ (i.e., the U-TAP is fully decomposable as there are no dependencies between the assignment decisions for the different tasks~\cite{chiang2007layering}). 

One possible solution to U-TAP is the \textbf{Uncapacitated Assignment (UA) Algorithm}~\ref{alg:UncapacitatedAssignment}, that exhaustively searches for the best execution option for each task. 
For each $j$, the UA runs a two-step procedure that first queries each $z \in Z$ for the best possible feasible $x \in X$ via the \textbf{Local Quality Maximizer} Procedure~\ref{alg:LQM}. It then selects $z$ that obtains the highest overall utility. The complexity of the UA is $\mathcal{O}(|J| \cdot |Z|\cdot |X|)$.

\subsection{Solving C-TAP \label{Sect:SolvingCapacitatedProblem}}

\par 
While the C-TAP is NP-hard, 
like other knapsack problems, for relatively small problem sizes it can be solved exactly~\cite{kellerer2004knapsack}. 

\par 
The following four-step \textbf{Assign-Tasks (AT) Algorithm} solves the C-TAP problem optimally. For simplicity, we show step 3 of this algorithm for the specific case of two capacitated nodes. We also demonstrate a simplification of this step for the case of one capacitated node.  

In the steps of this algorithm outlined below, the set of nodes $Z$ consists of subsets $Z^{\infty}$ and $Z^{cap}$, denoting, correspondingly, infinite-capacity and finite-capacity nodes.

\begin{algorithm}[t]
\caption{Complete Uncapacitated Assignments (CUA)}\label{alg:CloudPlacements}
\begin{algorithmic}[1]
\Procedure{CUA}{$J, Z$}
\State $J^{pld} \leftarrow \emptyset$
\For{$j = 1; j \leq |J|; j++$} 
    \State $\{z^{best}, x^{best}, u^{best}\} \leftarrow UA(j,Z)$
    \If{$z^{best} \in Z^{\infty}$}
    	\State $\{z_j^{*}, x_j^{*}, u_j^{*}\}\leftarrow \{z_j^{best}, x_j^{best}, u_j^{best}\}$ 
        \State $J^{pld} \leftarrow J^{pld} \cup j$
    \EndIf
\EndFor
\State \textbf{return} $\{z_j^{*}, x_j^{*}, u_j^{*}\}$ \ $ \forall j \in J^{pld}$, $J' \leftarrow J \setminus J^{pld}$
\EndProcedure
\end{algorithmic}
\end{algorithm}

\par First, via the \textbf{Complete Uncapacitated Assignments} Procedure~\ref{alg:CloudPlacements}, we identify the task set $J^{pld}$ of the tasks that achieve best performance when executed on the infinite-capacity nodes $z \in Z^{\infty}$. These tasks do not ``compete'' for the space on the capacitated nodes, and hence can be removed from the subsequent calculations. Following this step, we operate on the reduced set of tasks $J' \leftarrow J  \setminus J^{pld}$. The complexity of this procedure is same as the complexity of the UA algorithm, $\mathcal{O}(|J| \cdot |Z|\cdot |X|)$.


\begin{algorithm}[t]
\caption{Calculate Capacitated Gains (CCG)}\label{alg:CapGains}
\begin{algorithmic}[1]
\Procedure{CCG}{$J', Z^{\infty}, Z^{cap}$}
\For{$j = 1; j \leq |J'|; j++$} 
    \State $\{z_j^{\infty}, x_j^{\infty}, u_j^{\infty}\} \leftarrow UA(j,Z^{\infty})$
\EndFor
\For{$z=1; z\leq |Z^{cap}|; z++$}
	\For{$j = 1; j \leq |J'|; j++$} 
	\State $\{x_j^{z}, u_j^{z}\} \leftarrow LQM_{z}(j, T_{jzx}, A_{jx})$
    \EndFor
\EndFor
\State $u^{cap}(j,z) \leftarrow u_j^{z} - u_{j}^{\infty}$
\State \textbf{return} $\{u^{cap}(j,z), z_j^{\infty}, x_j^{\infty}, u_j^{\infty}\} $
\EndProcedure
\end{algorithmic}
\end{algorithm}

\par 
Then, via the \textbf{Calculate Capacitated Gains} Procedure~\ref{alg:CapGains}, we calculate the value of assigning tasks to each of the capacitated nodes, in comparison to assigning them to the uncapacitated ones. In this procedure, for each task $j \in J'$, we calculate the UA over the set of uncapacitated nodes -- the operation of total complexity $\mathcal{O}(|J'| \cdot |Z^{\infty}| \cdot |X|)$. Then, for each $j\in J'$, for all nodes in $Z^{cap}$, we calculate the achievable utility, by going over all $X$ on each node $z$. The complexity of this operation is  $\mathcal{O}(|J'| \cdot |Z^{cap}| \cdot |X|)$. 

\begin{algorithm}[t]
\caption{Choose Tasks for Capacitated Nodes (CTC) }\label{alg:DynProg}
\begin{algorithmic}[1]
\Procedure{CTC}{$J'$, $u^{cap}(j,z)$ $\forall$ $j\in J', z \in Z^{cap}$}
\State $jSet_{z1}(j,c1,c2) \leftarrow \emptyset$, $jSet_{z2}(j,c1,c2) \leftarrow \emptyset$ \Comment{Instantiate sets of tasks to be assigned to $Z_1$ and $Z_2$}
\State $h(j,c_1,c_2) \leftarrow -\infty$ \ \ $\forall$ $j \leq |J'|, c_1 \leq C_1, c_2 \leq C_2$
\For{$j=1; j \leq |J'|; j++$}
	\For{$c_1=0; c_1 \leq C1; c_1++$}
    	\For{$c_2=0; c_2 \leq C2; c_2++$}
        	\State $h(j,c_1,c_2) \leftarrow \max [h(j-1,c_1-1,c_2)+u^{cap}(j,1), h(j-1,c_1,c_2-1)+u^{cap}(j,2), h(j-1,c_1,c_2)]$
		    \If{$h(j,c_1,c_2)==h(j-1,c_1-1,c_2)+u^{cap}(j,1)$} \Comment{Task should be assigned to $Z_1$}
                \State $jSet_{z1}(j,c1,c2) \leftarrow jSet_{z1}(j-1,c1-1,c_2) \cup j$
		    \ElsIf{$h(j,c_1,c_2)==h(j-1,c_1,c_2-1)+u^{cap}(j,2)$} \Comment{Task should be assigned to $Z_2$}
                \State $jSet_{z2}(j, c1,c2) \leftarrow jSet_{z2}(j-1, c1,c2-1) \cup j$
            \ElsIf{$h(j,c_1,c_2)==h(j-1,c_1,c_2)$} \Comment{Task should not be assigned to either capacitated node}
                \State $J^{unp} \leftarrow J^{unp} \cup j$
    		\EndIf 
\EndFor
\EndFor
\EndFor
\State \textbf{return} $\underset{0 \leq c_1 \leq C_1, 0 \leq c_2 \leq C_2}{\max} h(|J'|, c_1, c_2)$, and the associated $jSet_{z1}(|J'|, c1, c2)$, $jSet_{z2}(|J'|,c1,c2)$, $J^{unp}$
\EndProcedure
\end{algorithmic}
\end{algorithm}

\par 
Then, in the \textbf{Choose Tasks for Capacitated Nodes} Procedure~\ref{alg:DynProg}, we use a conventional bottom-up dynamic programming approach to select the collection of tasks that need to be executed on each of the capacitated nodes. 
For simplicity, we show this procedure for the case of two capacitated nodes, $Z_1, Z_2$, with capacities $C_1$, $C_2$; it can be easily extended to include additional capacitated nodes. 
In this procedure, iterating over tasks $j$ and capacitated nodes' ``fill levels'' $c_1$, $c_2$, we calculate the value of a state $(j,c_1,c_2)$, as 
\begin{align} h(j,c_1,c_2)  \leftarrow \max & [
 h(j-1,c_1-1,c_2)+u^{cap}(j,1), \\ \notag 
& h(j-1,c_1,c_2-1)+u^{cap}(j,2), \\
& h(j-1,c_1,c_2)], \notag
\end{align}
where, within the maximization operation, the first term corresponds to assigning task $j$ to node $Z_1$, the second term -- to assigning it to node $Z_2$, and the third -- to not accepting the task. 
The solution that maximizes $\underset{0 \leq c_1 \leq C_1, 0 \leq c_2 \leq C_2}{\max} h(|J'|, c_1, c_2)$ is the optimal (if $|J|\geq C_1+C_2$, the maximum is obtained at $h(|J'|, C_1, C_2)$). 
The complexity of this procedure is $\mathcal{O}(|J'| \cdot |C_1| \cdot |C_2|)$, which is reasonable in practice for relatively small $C_1$ and $C_2$. 

\begin{algorithm}[t]
\caption{Reject Unassignable Tasks}\label{alg:RejectUnplaceable}
\begin{algorithmic}[1]
\Procedure{RUT}{$J^{unp}$, $\{z_j^{\infty}, x_j^{\infty}, u_j^{\infty}\}$ \ $\forall j \in J^{unp}$}
\For{$j = 1; j \leq |J^{unp}|; j++$} 
    \If{$u_j^{\infty} > 0$}
    	\State $\{z_j^{*}, x_j^{*}, u_j^{*}\} \leftarrow \{z_j^{\infty}, x_j^{\infty}, u_j^{\infty}\}$, $J^{pld} \leftarrow J^{pld}\cup j$
	\EndIf
\EndFor
\State \textbf{return} $\{z_j^{*}, x_j^{*}, u_j^{*}\}$ \ $ \forall \ j \in J^{pld}$, 
$J^{unp} \leftarrow J \setminus J^{pld}$
\EndProcedure
\end{algorithmic}
\end{algorithm}

\par 
Finally, via the \textbf{Reject Unassignable Tasks} Procedure~\ref{alg:RejectUnplaceable}, we identify the tasks that can be assigned to one of the infinite-capacity nodes. 
The remaining tasks, that cannot be assigned to infinite-capacity nodes and that were not selected for execution on the finite-capacity ones, are rejected by the system. The complexity of this procedure is $\mathcal{O}|J^{unp}|$.  
For the case of a single capacitated node, e.g., one capacitated gateway with capacity $C_1$, step 3 simplifies further to picking the $C_{1}$ tasks with the highest capacitated gains. This solution is optimal as in this case our problem is equivalent to a single knapsack problem with all weights of one, which is trivially solved by selecting the highest-value items.

\subsection{Numerical results~\label{Sect:SimulationResults}}

We implemented the developed algorithms in MATLAB, and used both synthetic and experimental data to validate them. 

The algorithms handle differentiated task timeliness preferences in the expected manner. 
For instance, we considered the settings where 10 tasks $j$ had different associated timeliness functions, of the wait-readily-first category, specifically $f_{j}(t) \leftarrow 1$ for $t \leq 0.3 \cdot j$, 
$f_j(t) \leftarrow 1-\frac{t-0.3}{j\cdot 0.1}$ for $0.3 \leq t \leq 0.3+j\cdot 0.1$, and $f_j(t) \leftarrow 0$ for $0.3+j \cdot 0.1 \leq t, \forall \ j \in J $. 
This set of timeliness functions corresponds to progressively loosening execution timeline preferences for tasks with increasing indexes -- e.g., for the first task, $f_1(t) \leftarrow 0$ for $t \geq 0.4$, while for the last task, $f_{10}(t) \leftarrow 0$ only for $t \geq 1.3$. We considered the settings with two nodes, a gateway node $1$ and a cloud node $2$, with intrinsic utilities $A_{j1}=0.6, A_{j2}=0.9 \ \forall j\in J$, $|X|=1$ (only one execution option on each of the nodes), and gateway and cloud execution latencies distributed uniformly, $T_{j11}\sim U[0.1, 0.6]$ s, $T_{j21} \sim U [0.3, 0.8]$ s, $\forall j \in J$. 
In this case, when both the gateway and the cloud are uncapacitated ($C_1= \infty, C_2 = \infty$), the UA algorithm assigns the first five tasks for execution on the gateway, and the next five tasks for execution on the cloud. As we add capacity restrictions to the gateway, for $C_1=1,2,$... the AT correctly assigns the most 1, 2, ... time-pressed tasks to the gateway for execution. 

The algorithms do more than simply place the least lax tasks on the nodes that offer the lowest latencies, however. We considered, for example, the settings described above, but with the intrinsic utility of cloud execution $A_{j2}$ randomly varying between 0.6 and 0.9 for the different tasks $j$. 
In this case, when the gateway is capacitated with $C_1=3$, over 10,000 executions of the AT algorithm with the randomizes $A_{j2}$ values, the algorithm assigns the $4^{\textrm{th}}$ lax task to be executed on the gateway in 32\% of cases, the $5^{\textrm{th}}$ lax task in 6.8\% of cases, and the $6^{\textrm{th}}$ lax task in 0.6\% of cases -- the algorithm co-optimizes the time and quality for the individual tasks, and takes into account relative benefits of gateway versus cloud processing for the different ones. 

As expected, the UA and the AT algorithms out-perform the approaches that minimize the latency or maximize the quality for all tasks. 
In the above-described settings, for example, a min-latency approach assigns all tasks to the gateway, while the max-quality approach assigns all tasks to the cloud. In this scenario, for $A_{j1}=0.6,A_{j2}=0.9 \ \forall j \in J$, the average utility achieved by the UA is 0.5078, while the average utility achieved by the min-latency approach is 0.4677, and the average utility achieved by the max-quality approach is 0.4415. This demonstrates that our algorithms correctly match tasks to execution locations that best fit task requirements. 

The algorithms also work well when multiple execution points are capacitated. 
For example, for nodes with intrinsic utilities $A_{j1} = 0.6, A_{j2} = 0.7, A_{j3} = 0.9 \ \forall \ j\in J$, capacities $C_1 = C_3=3, C_2 = \infty$, and task execution latencies $T_{j11}\sim U[0.5, 0.7]$~s, $T_{j12}\sim U [1.3, 1.5]$~s, and $T_{j13} \sim U [1.6, 1.9]$~s (i.e., one capacitated node offers quick low-quality service, while another offers slow service of higher quality),  for $f_{j}(t) \leftarrow 1$ for $t \leq 0.5 \cdot j$, 
$f_j(t) \leftarrow 1-\frac{t-0.5}{j\cdot 0.3}$ for $0.5 \leq t \leq 0.5+j\cdot 0.3$, and $f_j(t) \leftarrow 0$ for $0.5+j \cdot 0.3 \leq t, \forall \ j \in J $, 
the C-TAP correctly assigns the most time-pressed 3 tasks for execution on node 1, 
and the 2 most lax tasks for execution on node 3. 

In the in-flight conditions we captured in our experiments (i.e., the conditions depicted in Fig.~\ref{fig:CDFSeattleFlights}), our algorithms assign different number of tasks to local nodes depending on the connectivity conditions. 
For example, we considered the case with task completion latencies distributed according to our experimental latency measurements for a local node for node $1$,  and our experimental measurements for the Region~1 AWS EC2 node for node $2$, for 100 tasks $j \in J $ with timeliness functions 
$f_{j}(t) \leftarrow 1$ for $t \leq 0.5$, 
$f_j(t) \leftarrow 1-\frac{t-0.5}{j\cdot 0.2}$ for $0.5 \leq t \leq 0.5+j\cdot 0.2$, 
and $f_j(t) \leftarrow 0$ for $0.5 +j\cdot 0.2 \leq t, A_{j1} = 0.6, A_{j2} = 0.9\ \forall j \in J$, 
and $|X|=1$.  
In this case, according to the UA algorithm, on the flight with the best connectivity (location 4 in Table~\ref{table:TripToSeattle}), all 100 tasks should be executed on the cloud, 
on the flight with the average connectivity (location 6 in Table~\ref{table:TripToSeattle}), 
22 most time-pressed tasks should be executed on a local node, 
and on the flight with the worst cloud connectivity (location 10 in Table~\ref{table:TripToSeattle}), 
56 most time-demanding tasks should be executed locally.  

\balance


\section{Conclusions \label{Sect:Conclusions}}

In this work we examined properties of task completion latencies in fog computing systems with multiple heterogeneous execution points. First, we conducted a benchmarking-based study of task completion latencies in fog computing architectures, which included 6 heterogeneous execution points. 
Using the developed setup, we completed over 1,000 hours of targeted experiments. 
This study elucidated properties of task completion latencies in fog computing that have not been examined before. 
Armed with the insights from the experimental study, we then developed a framework to co-optimize task completion latency and quality in task assignments in fog computing. We developed task assignment problem formulations that take into account important properties of fog computing systems that were not examined before, such as potential presence of execution points with infinite capacities, and developed algorithms for solving them. In future work we will validate our algorithms via deploying them in realistic fog conditions, and will our measurements publicly available.

\balance

\section{Acknowledgements}

This work was supported in part by the Comcast Innovation Fund Research Grant, AWS Cloud Credits for Research, Microsoft Azure Research Award, NSF CSR-1812797 grant, and Defense Advanced Research Projects Agency (DARPA) under contract No. HR001117C0052 and No. HR001117C0048. The opinions, findings and conclusions expressed in this material are those of the author(s) and do not necessarily reflect the views of the Defense Advanced Research Projects Agency.
\balance
\newpage 
\bibliographystyle{acm}
\bibliography{ResearchStatement2016}
\end{document}